\documentclass[fleqn,usenatbib]{mnras}

\usepackage{newtxtext,newtxmath}

\usepackage[T1]{fontenc}

\DeclareRobustCommand{\VAN}[3]{#2}
\let\VANthebibliography\thebibliography
\def\thebibliography{\DeclareRobustCommand{\VAN}[3]{##3}\VANthebibliography}

\usepackage{graphicx}    
\usepackage{amsmath}    
\usepackage{xcolor}

\newcommand{\hhh}{$^{\mathrm{h}}$}
\newcommand{\mmm}{$^{\mathrm{m}}$}
\newcommand{\sss}{$^{\mathrm{s}}$}
\newcommand{\ddd}{$^{\mathrm{\circ}}$}
\newcommand{\dmm}{$^{\prime}$}

\newcommand{\HI}{$\textrm{H}\scriptstyle\mathrm{I}$}

\title[MIGHTEE-HI: the COSMOS field]{MIGHTEE-\HI: deep spectral line observations of the COSMOS field}

\author[Heywood et al.]
{\parbox{\textwidth}{
\begin{flushleft}
I.~Heywood$^{1,2,3}$\thanks{E-mail: ian.heywood@physics.ox.ac.uk}, 
A.~A.~Ponomareva$^{1}$, 
N.~Maddox$^{4}$, 
M.~J.~Jarvis$^{1,5}$, 
B.~S.~Frank$^{6,7,3,8}$, 
E.~A.~K.~Adams$^{9,10}$, 
M.~Baes$^{11}$, 
A.~Bianchetti$^{12,13}$, 
J.~D.~Collier$^{14,8}$,
R.~P.~Deane$^{15,16}$, 
M.~Glowacki$^{17}$, 
S.~L.~Jung$^{1}$,
H.~Pan$^{1}$, 
S.~H.~A.~Rajohnson$^{7}$, 
G.~Rodighiero$^{12,13}$, 
I.~Ruffa$^{18,19}$, 
M.~G.~Santos$^{5,3}$, 
F.~Sinigaglia$^{20,21}$, 
and M.~Vaccari$^{8,18}$
\\
\end{flushleft}
}
\\
\footnotesize
\\
$^{1}$Astrophysics, Department of Physics, University of Oxford, Keble Road, Oxford, OX1 3RH, UK\\ 
$^{2}$Centre for Radio Astronomy Techniques and Technologies, Department of Physics and Electronics, Rhodes University, PO Box 94, Makhanda 6140, South Africa\\
$^{3}$South African Radio Astronomy Observatory, 2 Fir Street, Black River Park, Observatory 7925, South Africa\\
$^{4}$School of Physics, H.H. Wills Physics Laboratory, Tyndall Avenue, University of Bristol, Bristol, BS8 1TL, UK\\
$^{5}$Department of Physics and Astronomy, University of the Western Cape, Robert Sobukwe Road, 7535 Bellville, Cape Town, South Africa\\
$^{6}$STFC UK Astronomy Technology Centre, Royal Observatory, Edinburgh, Blackford Hill, Edinburgh, EH9 3HJ\\
$^{7}$Department of Astronomy, University of Cape Town, Private Bag X3, Rondebosch 7701, South Africa\\
$^{8}$The Inter-University Institute for Data Intensive Astronomy (IDIA), and University of Cape Town, Private Bag X3, Rondebosch, 7701, South Africa\\
$^{9}$ASTRON, the Netherlands Institute for Radio Astronomy, Oude Hoogeveesedĳk 4,7991 PD Dwingeloo, The Netherlands\\
$^{10}$Kapteyn Astronomical Institute, University of Groningen, PO Box 800, 9700 AV Groningen, The Netherlands\\
$^{11}$Sterrenkundig Observatorium, Universiteit Gent, Krĳgslaan 281 S9, B-9000 Gent, Belgium\\
$^{12}$Department of Physics and Astronomy, Università degli Studi di Padova, Vicolo dell’Osservatorio 3, I-35122, Padova, Italy\\
$^{13}$INAF - Osservatorio Astronomico di Padova, Vicolo dell’Osservatorio 5, I-35122, Padova, Italy\\
$^{14}$Curtin Institute for Radio Astronomy, 1 Turner Ave, Bentley WA 6102, Australia\\
$^{15}$Wits Centre for Astrophysics, University of the Witwatersrand, 1 Jan Smuts Avenue, 2000, Johannesburg, South Africa\\
$^{16}$Department of Physics, University of Pretoria, Hatfield, Pretoria, 0028, South Africa \\
$^{17}$International Centre for Radio Astronomy Research (ICRAR), Curtin University, Bentley, WA 6102, Australia\\
$^{18}$Cardiff Hub for Astrophysics Research \&\ Technology, School of Physics \&\ Astronomy, Cardiff University, Queens Buildings, The Parade, Cardiff, CF24 3AA, UK\\
$^{19}$INAF-Istituto di Radioastronomia, via Gobetti 101, I-40129 Bologna, Italy\\
$^{20}$Département d’Astronomie, Université de Genève, Chemin Pegasi 51, 1290 Versoix, Switzerland\\
$^{21}$Institut für Astrophysik, Universität Zürich, Winterthurerstrasse 190, CH-8057 Zürich, Switzerland\\
}

\date{Accepted 2024 September 03. Received 2024 September 03; in original form 2024 March 19}

\pubyear{2024}

\begin{document}
\label{firstpage}
\pagerange{\pageref{firstpage}--\pageref{lastpage}}
\maketitle

\begin{abstract} 

\noindent
The MIGHTEE survey utilises the South African MeerKAT radio telescope to observe four extragalactic deep fields, with the aim of advancing our understanding of the formation and evolution of galaxies across cosmic time. MIGHTEE's frequency coverage encompasses the \HI~line to a redshift of z~~$\simeq$~0.58, and OH megamasers to z~$\simeq$~0.9. We present the MIGHTEE-\HI~imaging products for the COSMOS field, using a total of 94.2~h on-target and a close-packed mosaic of 15 individual pointings. The spectral imaging covers two broad, relatively interference-free regions (960--1150 and 1290--1520~MHz) within MeerKAT's L-band, with up to 26~kHz spectral resolution (5.5 km~s$^{-1}$ at $z$~=~0). The median noise in the highest spectral resolution data is 74 $\mu$Jy beam$^{-1}$, corresponding to a 5$\sigma$ \HI~mass limit of 10$^{8.5}$~M$_{\odot}$ for a 300 km~s$^{-1}$ line at $z$~=~0.07. The mosaics cover $>$4~deg$^{2}$, provided at multiple angular resolution / sensitivity pairings, with an angular resolution for \HI~at $z$~=~0 of 12$''$. We describe the spectral line processing workflow that will be the basis for future MIGHTEE-\HI~products, and validation of, and some early results from, the spectral imaging of the COSMOS field. We find no evidence for line emission at the position of the $z$~=~0.376 \HI~line reported from the CHILES survey at a $>$94 per cent confidence level, placing a 3$\sigma$ upper limit of 8.1~$\times$~10$^{9}$~M$_{\odot}$ on $M_{\mathrm{HI}}$ for this galaxy. A public data release accompanies this article.
 
\end{abstract}




\section{Introduction}
\label{sec:intro}

Neutral hydrogen (\HI) is the most abundant element in the Universe. It serves as the fundamental fuel reservoir for star formation in galaxies \citep[e.g.][]{huang2012,maddox2015}, as well as the growth of their central supermassive black holes (SMBH), and the fuelling of activity associated with them \citep[e.g.][]{maccagni2023}. Understanding the role of \HI~in various galactic processes is therefore also key to understanding the formation and evolution of galaxies as a whole, and obtaining a picture of this process over cosmic time remains one of the major goals of modern astrophysics. 

The 1420~MHz (21~cm; rest-frame) spectral line emitted by \HI~\citep{ewen1951} is the most powerful observational tracer available to us in pursuit of understanding the role of atomic hydrogen. Many large-scale observational campaigns targeting this line with many single-dish and synthesis array radio telescopes have been conducted over the years. These include large-area surveys that cover up to all of the sky visible to an observatory, primarily targeting the line at low redshift, such as the \HI~Parkes All-Sky Survey \citep[HIPASS;][]{barnes2001}, the Arecibo Legacy Fast ALFA survey \citep[ALFALFA;][]{giovanelli2005}, the Effelsberg-Bonn \HI~Survey \citep[EBHIS;][]{kerp2011}, and the imaging survey conducted with the Aperture Tile In-Focus (APERTIF) upgrade to the Westerbork Synthesis Radio Telescope \citep[WSRT;][]{adams2022}. These are complemented by narrower, deeper surveys that aim to detect \HI~at higher redshifts, for example the Blind Ultra Deep \HI~Environmental Survey \citep[BUDHIES;][]{verheijen2007}, the COSMOS \HI~Large Extragalactic Survey \citep[CHILES;][]{fernandez2013}, and the HIGHz survey \citep{catinella2015}. A major reason for targeting \HI~at high redshift is not only to study individual galaxies and ensembles of galaxies in varying environments, but also to obtain measurements of the cosmic evolution of the \HI~density of the universe ($\Omega_{\mathrm{HI}}$). It is well established that the star formation rate density of the Universe peaks at $z$~$\sim$~1--3 (`cosmic noon') in tandem with the growth of SMBHs, and declines to the present day \citep{madau2014,aird2018,walter2020}, however present observational constraints on $\Omega_{\mathrm{HI}}$ show comparatively weak evolution over the same epochs \citep[e.g.][]{hu2019,grasha2020,chen2021}. A complete census of cold gas, including both the atomic and molecular phases is needed to resolve this. Existing and on-going efforts to study the cosmic evolution of molecular gas \citep[e.g.][]{decarli2019,audibert2022} motivates the need for large volume, untargeted surveys that are capable of detecting large, unbiased samples of galaxies via their \HI~line, out to high redshifts.

The detection of \HI~at cosmologically significant distances is challenging due to the intrinsic faintness of the \HI~line. Indeed, the telescope collecting area required to achieve a direct detection at cosmic noon \citep{wilkinson1991} was what gave the forthcoming Square Kilometre Array (SKA) observatory its name, and observations of the \HI~line have remained a primary scientific driver for this facility as its design has evolved over the last three decades. The SKA precursor facilities thus also naturally have \HI~as a primary scientific driver that has influenced both the design of the telescopes and their flagship observational programmes. The South African MeerKAT telescope \citep{jonas2016} is one such instrument, consisting of 64~$\times$~13.5~m offset-Gregorian antennas, with a dense core (70 per cent of the collecting area within 1~km radius), and a maximum antenna separation of 8~km. The antenna design provides a wide field of view, and this along with a trio of low system temperature receivers, and the aforementioned antenna layout make MeerKAT an extremely fast survey telescope, with interferometric imaging fidelity that is unprecedented at these wavelengths \citep[e.g.][]{condon2021,knowles2022,heywood2022b}.

The large survey projects that MeerKAT is undertaking include studies of \HI~at very low column densities in nearby galaxies \citep[MeerKAT \HI~Observations of Nearby Galactic Objects: Observing. Southern Emitters, or MHONGHOOSE;][]{deblok2024} and cluster environments via the Fornax survey \citep{serra2023}, and projects that target \HI~out to $z$~$\sim$~1 through the MeerKAT Absorption Line Survey \citep[MALS;][]{gupta2021}, as well as via direct detections and statistical methods in the ultra-deep, single pointing observations of the Looking At the Distant Universe with the MeerKAT Array survey \citep[LADUMA;][]{blyth2016}. 

The MeerKAT GHz Tiered Extragalactic Explorations (MIGHTEE) survey \citep{jarvis2016} falls into the latter category. It has amassed $\sim$1000 hours of MeerKAT observations using the L-band receivers (856--1712 MHz). The overarching goal of the survey is to determine how galaxies form and evolve as a function of cosmic time and environment. The aim is to achieve this by exploiting the full dimensionality of the data that MeerKAT provides. The data products for the survey encompass total intensity continuum \citep{heywood2022a}, spectro-polarimetry \citep{taylor2024}, and high resolution (26 kHz; $\Delta v$~=~5.5~km~s$^{-1}$ at $z$~=~0) spectral line imaging \citep{maddox2021}, the primary goal of the latter being the detection of redshifted \HI~emission, as well as line emission from hydroxyl (OH) megamasers via the hyperfine transitions at 1612, 1665, 1667 and 1720~MHz. The survey also targets four of the most data-rich extragalactic deep fields, namely the COSMOS field, the Chandra Deep Field South (CDFS), the southern region of the European Large Area ISO Survey (ELAIS-S1), and the XMM-\emph{Newton} Large Scale Structure field (XMM-LSS).

An initial set of Early Science spectral line products \citep{maddox2021} have been produced using MIGHTEE data, consisting of a single pointing in the COSMOS field (16 h on-source), and three (non-mosaicked) pointings in the XMM-LSS field (each with 12 h on-source). These made use of the initial set of MIGHTEE observations where the MeerKAT correlator was configured to only deliver 4096 channels across the full L-band. Despite the reduced spectral resolution (208.8 kHz; $\Delta v$~=~44~km~s$^{-1}$ at $z$~=~0) this initial set of data has yielded numerous results, including the discovery of a new galaxy group \citep{ranchod2021}, evidence of tidal interactions in nearby galaxies \citep{namumba2023}, studies of the relationship between the \HI~gas in galaxies and the cosmic web \citep{tudorache2022}, an investigation of the average \HI~mass of galaxies as a function of their environment \citep{sinigaglia2024}, and pushed the redshift envelope for measurements of the \HI~mass function \citep{ponomareva2023} and several fundamental \HI~scaling relationships \citep{ponomareva2021,rajohnson2022,sinigaglia2022,pan2023}.

In this paper we present the first data release (Data Release 1, hereafter DR1) for the spectral line component of the MIGHTEE survey (MIGHTEE-\HI), and the first to feature images at the full spectral resolution of the data (26~kHz, corresponding to 5.5 km~s$^{-1}$ at $z$~=~0). We make use of all of the 32,768 channel observations that have targetted the COSMOS field, to produce mosaicked spectral line cubes covering $\sim$4~deg$^{2}$. 

In this paper we adopt $H_{0}$~=~67.4~km~s$^{-1}$, $\Omega_{\mathrm{m}}$~=~0.315, and $\Omega_{\mathrm{\Lambda}}$~=~0.685 \citep{planck2020} as the $\mathrm{\Lambda}$CDM cosmological parameters for any calculations that require them.

\section{Observations}
\label{sec:obs}

\begin{figure*}
\centering
\includegraphics[width=0.9 \textwidth]{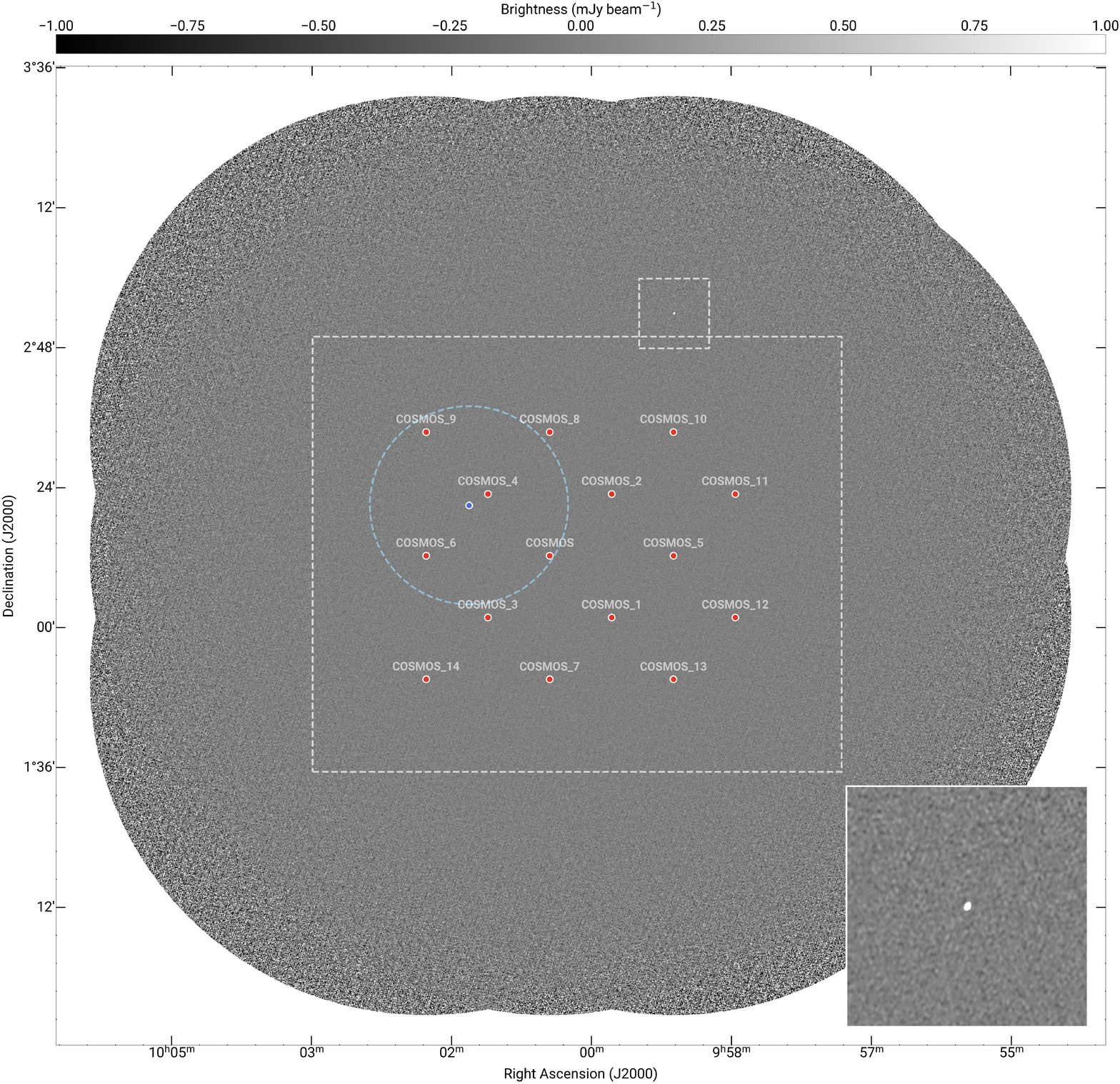}
\caption{An overview of the COSMOS field, showing the centre locations of the fifteen MeerKAT pointings that we make use of for this work (see Table \ref{tab:obs} for coordinates). The large dashed rectangle shows the coverage of the ultra-deep \emph{YJHK$_{s}$} near-infrared imaging from UltraVISTA \citep{mccracken2012}. The coverage of the MeerKAT imaging is entirely within the ultra-deep \emph{grizy} band imaging from the Subaru Hyper Suprime-Cam \citep{aihara2018,aihara2019}. The blue marker and corresponding concentric dashed circle respectively show the pointing centre and aereal coverage of the CHILES deep \HI~survey \citep{fernandez2013}. The background image (TAN projection) is a single channel at 975.5~MHz from the MIGHTEE-\HI~data that we present in this article. The small dashed square and corresponding inset image show the peak line emission from one of the first discoveries from these data products, namely the highest redshift ($z$~=~0.7012) hydroxyl megamaser discovered to date \citep{jarvis2024}.}
\label{fig:schematic}
\end{figure*}

The MIGHTEE survey has acquired 992~h of observations with MeerKAT's L-band system, of which 805.3~h are observations of science targets, with the remainder of the time being used for scans of calibrator sources. Observations that employ the full spectral resolution mode of the correlator, having 32,768 frequency channels, form 77 per cent of these. The full spectral resolution data breakdown per field is: CDFS 36 per cent; COSMOS 16 per cent; ELAIS-S1 10 per cent; XMM-LSS 38 per cent. The initial 23 per cent of the MIGHTEE data was taken prior to the deployment of the full spectral resolution mode of the MeerKAT correlator, and was recorded with 4,096~$\times$~208~kHz channels. An initial `Early Science' pilot data set for MIGHTEE-\HI~was produced from these data, featuring three pointing centres in XMM-LSS and one in COSMOS \citep[see e.g.][]{ponomareva2021}. This initial, reduced resolution data set formed the basis of the numerous \HI~studies already mentioned in Section \ref{sec:intro}.

In this paper we present the full spectral resolution data for the COSMOS field in their entirety. The data consist of 15 pointings of approximately 8~h each, for a total of 120~h. The science target field observations made up 94.2~h of this total, with the remainder of the time used for calibration purposes. The pointing centres were arranged in a close packed mosaic (as shown in Figure \ref{fig:schematic}), with separations of 21$'$ and 11$'$ in right ascension and declination respectively (corresponding to 35 and 18 per cent of the half power primary beam level at 1420 MHz). Driven by the sky coverage of the multiwavelength data, the mosaic pattern deployed for the COSMOS field delivers increased sensitivity over the pointing strategy used for the other MIGHTEE fields, which is (largely) a standard hexagonal pattern critically spaced at the upper frequency of L-band. The original goal for this pattern was to deliver a thermal noise level of 100~$\mu$Jy beam$^{-1}$ in a 26~kHz channel \citep{maddox2021}. The sensitivity of the final image products is explored in detail in Section \ref{sec:sensitivity}. Further details of the observations and a list of their pointing centres can be found in Table \ref{tab:obs}. As stated above, for each observation the correlator was configured to deliver 32,768 frequency channels, and 8 s correlator integration times. The primary and secondary calibrators were J0408$-$6545 and J1008+0740 respectively. The primary calibrator was visited for $\sim$10 minutes, twice per observation, and the secondary was visited for 2--3 minutes for every 20--30 minutes of target observing.

\begin{table*}
\begin{minipage}{176mm}
\centering
\caption{List of the MeerKAT pointings used in this work, in chronological order of observation date. The final row shows the total observing time, the total on-source time, and the time-weighted average number of MeerKAT antennas used for this project.}
\begin{tabular}{lllllrrr} \hline
Date       & ID         & Field  & RA          & Dec        & Track   & On-source & N$_{\mathrm{ant}}$  \\ 
           &            &        & (J2000)     & (J2000)    & (h)     & (h)       &                     \\ \hline
2020-03-28     & 1585413022     & COSMOS\_5       & 09\hhh59\mmm04\sss        & 02\ddd12\dmm21\farcs0     & 8              & 6.25           & 59              \\
2020-03-29     & 1585498873     & COSMOS\_6       & 10\hhh01\mmm54\sss        & 02\ddd12\dmm21\farcs0     & 8              & 6.25           & 59              \\
2020-03-31     & 1585671638     & COSMOS\_7       & 10\hhh00\mmm29\sss        & 01\ddd51\dmm08\farcs2     & 8              & 6.25           & 60              \\
2020-04-02     & 1585844155     & COSMOS\_8       & 10\hhh00\mmm29\sss        & 02\ddd33\dmm33\farcs8     & 8              & 6.25           & 60              \\
2020-04-30     & 1585928757     & COSMOS\_9       & 10\hhh01\mmm54\sss        & 02\ddd33\dmm33\farcs8     & 8              & 6.25           & 60              \\
2020-04-04     & 1586016787     & COSMOS\_10      & 09\hhh59\mmm04\sss        & 02\ddd33\dmm33\farcs8     & 8.03           & 6.25           & 60              \\
2020-04-06     & 1586188138     & COSMOS\_11      & 09\hhh58\mmm21\sss        & 02\ddd22\dmm57\farcs4     & 8              & 6.25           & 59              \\
2020-04-07     & 1586274966     & COSMOS\_12      & 09\hhh58\mmm21\sss        & 02\ddd01\dmm44\farcs6     & 8              & 6.25           & 60              \\
2020-04-12     & 1586705155     & COSMOS\_13      & 09\hhh59\mmm04\sss        & 01\ddd51\dmm08\farcs2     & 8              & 6.25           & 59              \\
2020-04-13     & 1586791316     & COSMOS\_14      & 10\hhh01\mmm53\sss        & 01\ddd51\dmm08\farcs2     & 8              & 6.25           & 60              \\
2020-04-26     & 1587911796     & COSMOS          & 10\hhh00\mmm29\sss        & 02\ddd12\dmm21\farcs0     & 7.98           & 6.25           & 59              \\
2021-04-07     & 1617809470     & COSMOS\_1       & 09\hhh59\mmm46\sss        & 02\ddd01\dmm44\farcs6     & 8              & 6.25           & 60              \\
2021-05-02     & 1619963656     & COSMOS\_2       & 09\hhh59\mmm46\sss        & 02\ddd22\dmm57\farcs4     & 7.96           & 6.25           & 62              \\
2021-05-15     & 1621083675     & COSMOS\_4       & 10\hhh01\mmm11\sss        & 02\ddd22\dmm57\farcs4     & 7.97           & 6.25           & 61              \\
2021-05-30     & 1622376680     & COSMOS\_3       & 10\hhh01\mmm11\sss        & 02\ddd01\dmm44\farcs6     & 8              & 6.7            & 61              \\ \hline
               &                &                 &                           &                           & {\bf 119.9}    & {\bf 94.2}     & {\bf 59.9}      \\      
\hline
\end{tabular}
\label{tab:obs}
\end{minipage}
\end{table*}

\section{Data processing and image products}
\label{sec:dataproc}

Each of the fifteen observations described in Section \ref{sec:obs} is processed and imaged in a consistent way, before being mosaicked together to form the final image cubes. A description of the steps involved follows in this section. All processing was done using the \emph{ilifu} cluster\footnote{\scriptsize\url{https://www.ilifu.ac.za/}} in Cape Town, making use of {\sc singularity} for software containerisation\footnote{A container that provides all packages used is available from the {\sc oxkat} repository \citep{heywood2020b}.} \citep{kurtzer2017}. Following the initial preparation of the data described in Section \ref{sec:initial_proc}, the self-calibration, visibility-domain continuum subtraction, and imaging steps described in Sections \ref{sec:selfcal} and \ref{sec:imaging} are applied independently to each of two sub-band data sets, for each of the 15 individual pointings. The lower and upper frequency ranges of these sub-bands (and the corresponding redshift range for \HI), hereafter referred to as L1 and L2, are 960--1150 MHz (0.23~$<$~$z_{\mathrm{HI}}$~$<$~0.48) and 1290--1520 MHz (0~$<$~$z_{\mathrm{HI}}$~$<$~0.1) respectively (see also Table \ref{tab:proc_params}). Note that the upper frequency of the L2 band extends beyond the rest-frame frequency of \HI~in order to search for redshifted emission from OH megamasers via the lines listed in Section \ref{sec:intro}. The image products for all pointings are then brought together for homogenisation (Section \ref{sec:homogenise}) and mosaicking (Section \ref{sec:mosaicking}), with the final per-sub-band cube then subjected to an image-domain continuum subtraction process as described in Section \ref{sec:imcontsub}. Table \ref{tab:proc_params} also lists the pertinent parameters of the various processes described below.

\subsection{Initial visibility processing}
\label{sec:initial_proc}

The starting point for the processing involves retrieval of the visibilities from the South African Radio Astronomy Observatory's (SARAO) data archive\footnote{\scriptsize\url{https://archive.sarao.ac.za}}, using the KAT Data Access Library ({\sc katdal\footnote{\scriptsize\url{https://github.com/ska-sa/katdal}}}) to convert the data from the MeerKAT Visibility Format into Measurement Set (MS) format \citep{kemball2000}. One aspect of the \HI~processing that differs from the other strands of MIGHTEE is that we extract the Level-1 calibrated visibilities from the archive. These have instrumental gain corrections pre-applied, as derived by the SARAO Science Data Processor\footnote{\scriptsize\url{https://skaafrica.atlassian.net/wiki/spaces/ESDKB/pages/338723406/}} (SDP) software, which is deployed on a compute cluster in the Karoo Array Processor Building, and operates automatically on all MeerKAT observations. The SDP calibration pipeline derives delay and bandpass solutions from the primary calibrator (J0408$-$6545 in all cases), and time-dependent complex gain solutions from the secondary calibrator (J1008+0740 in all cases). Since a primary polarisation calibrator (3C 48 or 3C 138) was also present for all MIGHTEE observations, the SDP pipeline will also determine and apply cross-hand delay solutions, however this should have negligible effect on the total intensity data products that we subsequently produce.

The Level 1 MS is split into the two L1 and L2 sub-bands for subsequent processing.  The {\sc casa} \citep{casa2022} {\sc mstransform} task was used for this process, during which the data were also Doppler corrected to a Barycentric reference frame. To facilitate the parallelisation of subsequent processing steps that are amenable to it, the sub-band output data were written as multi-Measurement-Sets (MMS), with an individual sub-MS for each of the target scans.\footnote{A `scan' in this context is a contiguous observation of a particular pointing position. Each observing block may contain several scans of the same target field, as the scan number will change when for example the telescope switches to a calibrator source.} This is particularly beneficial for the flagging, self-calibration, and model visibility prediction steps. 

The first operation following production of the sub-band Measurement Set is a round of flagging using the {\sc tricolour} package \citep{hugo2022}. This software provides (amongst other functionality) an implementation of the {\sc sumthreshold} algorithm \citep{offringa2010}, accelerated using {\sc dask-ms} \citep{perkins2022} and optimised for MeerKAT.

\begin{table}
\centering
\caption{Sub-band properties and relevant processing parameters. Please refer to the relevant sections for details, as listed in the table. Note that properties of the data that are strongly frequency dependent (angular resolution and sensitivity) are addressed in detail in Sections \ref{sec:resolution} and \ref{sec:sensitivity}. Note that the redshift ranges for the OH molecule are calculated assuming $\nu_{\mathrm{rest}}$~=~1665 MHz.}
\begin{tabular}{lrr} \hline 
Sub-band                                 &    \bf{L1}     &   \bf{L2}    \\ \hline

\multicolumn{3}{c}{~}\\
\multicolumn{3}{c}{{\bf Frequency parameters} (Section \ref{sec:initial_proc})}\\
\multicolumn{3}{c}{~}\\
Lower frequency (MHz)                    &    960     &  1290    \\
Upper frequency (MHz)                    &   1150     &  1520    \\
Bandwidth (MHz)                          &    190     &   230    \\
$z_{\mathrm{min}}$ (\HI)                 &   0.23     &     0    \\
$z_{\mathrm{max}}$ (\HI)                 &   0.48     &   0.1    \\
$z_{\mathrm{min}}$ (OH)                  &   0.45     &   0.1    \\
$z_{\mathrm{max}}$ (OH)                  &   0.73     &  0.29    \\
Channel resolution (kHz)                 &  104.5     &  26.1    \\
Velocity resolution (z~=~0; km s$^{-1}$) &   22.1     &   5.5    \\
Number of channels                       &   1818     &  8805    \\ 

\multicolumn{3}{c}{~}\\
\multicolumn{3}{c}{{\bf Selfcal and visibility-domain continuum subtraction} (Section \ref{sec:selfcal})}\\
\multicolumn{3}{c}{~}\\
Image size (pixels)                      &  10240$^{2}$ & 10240$^{2}$\\
Pixel scale ($''$)                       &    1.1     &   1.1    \\
Robust weighting ($r$)                   & $-$0.3     & $-$0.3  \\
Clean threshold ($\mathrm{\mu}$Jy beam$^{-1})$  &      5     &     5   \\
Deconvolution (gridding) sub-bands       &     10     &    10    \\
Model (degridding) sub-bands             &    909     &  1761   \\
Model frequency polynomial order         &      4     &     4   \\
Selfcal solution interval (s)            &     32     &    32   \\

\multicolumn{3}{c}{~}\\
\multicolumn{3}{c}{{\bf Spectral imaging} (Sections \ref{sec:imaging} and \ref{sec:mosaicking})}\\
\multicolumn{3}{c}{~}\\
Image size (pixels; $r$~=~0.0, 0.5)                &  4096$^{2}$ & 4096$^{2}$ \\
Image size (pixels; $r$~=~1.0)                     &  1024$^{2}$ & 1024$^{2}$ \\
Pixel scale ($r$~=~0.0, 0.5; $''$)                 &    2.0      &   2.0       \\
Pixel scale ($r$~=~1.0; $''$)                      &    8.0      &   8.0       \\
Mask making threshold ($\sigma^{-1}$)              &    4.8      &   4.8        \\
Spatial mask dilation iterations                   &      5      &     5       \\
Spectral mask dilation iterations                  &      2      &     2         \\
Primary beam gain cut                              &    0.3      &   0.2       \\
Field of view (lowest frequency; deg$^{2}$)        &   6.49      &   4.12     \\
Field of view (highest frequency; deg$^{2}$)       &   4.94      &   4.08     \\
Band centre angular resolution ($r$~=~0.0; $''$)   & 14.8        & 11.7         \\
Band centre angular resolution ($r$~=~0.5; $''$)   & 19.6        & 15.6         \\
Band centre angular resolution ($r$~=~1.0; $''$)   & N/A         & 74.0         \\

\multicolumn{3}{c}{~}\\
\multicolumn{3}{c}{{\bf Image-domain continuum subtraction} (Section \ref{sec:imcontsub})}\\
\multicolumn{3}{c}{~}\\
Sigma clipping threshold ($\sigma^{-1}$) &    2.3     &   2.3   \\
Mask dilation iterations (channels)      &      4     &    16   \\
Median filter window (channels)          &      9     &    31   \\
Savitzky-Golay filter window (channels)  &      9     &    31   \\ 
\multicolumn{3}{c}{~}\\& &   \\ \hline
\end{tabular} 
\label{tab:proc_params}
\end{table}

\subsection{Self-calibration and visibility-domain continuum subtraction}
\label{sec:selfcal}

The starting point for the self-calibration and first stage of continuum subtraction of the full spectral resolution data for a particular pointing is its corresponding deep image created as part of MIGHTEE's total intensity continuum processing. These image products were subjected to direction-dependent calibration techniques and deep constrained deconvolution \citep[full details of the processing are provided by][]{heywood2022a}, and contain very few artefacts. They can thus provide the locations of radio sources down to flux densities that far exceed our depth requirements for the spectral line processing in each sub-band. We use the {\sc breizorro} tool \citep{ramaila2023} with the threshold parameter set to 5$\sigma$ to construct a deconvolution mask, which is visually inspected prior to proceeding. Each sub-band for each pointing is then deconvolved using {\sc wsclean} \citep{offringa2014}, using the relevant mask. For this process we must use the same image size and cell size parameters as the MIGHTEE continuum image, as listed in Table \ref{tab:proc_params} which also lists the clean termination thresholds and number of sub-bands over which the deconvolution takes place.

Following inspection of the residual and model images, the clean component models (with either 10 or 6 frequency planes, depending on the sub-band) are interpolated in frequency onto the higher number of frequency intervals listed in Table \ref{tab:proc_params} in order to impart spectral smoothness to the model. The {\sc smops}\footnote{\scriptsize{\url{https://github.com/Mulan-94/smops}}} tool is used for this, and the resulting set of frequency-dependent model images is then inverted into a corresponding set of model visibilities using {\sc wsclean} in predict mode.

A round of phase and delay self-calibration is applied to the data using the {\sc cubical} package \citep{kenyon2018}, with a time and frequency solution interval of 32 s and the entire width of the sub-band respectively. Visibility-domain continuum subtraction also takes place at this point, by having {\sc cubical} write the corrected residuals back to the MS, which involves applying the best fitting gain solutions to the data and subtracting the high spectral resolution sky model. The corrected residuals are then subjected to another round of flagging using {\sc tricolour}. A detailed examination of the flagging levels is presented in Appendix \ref{sec:flags}.

 Note that other MIGHTEE fields contain stronger, more problematic sources than the COSMOS field, and these may require subtraction via direction-dependent calibration schemes. Such a step can be implemented in the workflow using {\sc cubical} at this stage\footnote{Note that for future MIGHTEE-\HI~and continuum processing {\sc cubical} will be replaced by its faster and more computationally-efficient successor {\sc quartical} \citep{kenyon2022}. This will be highly beneficial for the spectral line processing given the scale of the data.}.

\subsection{Spectral imaging}
\label{sec:imaging}

The corrected residual data for each pointing and each sub-band must now be imaged on a per-channel basis. The first step is the straighforward inversion of the visibilities to produce dirty images for each channel. This is done using {\sc wsclean}, with four instances running in parallel on the L2 sub-band due to the larger number of channels. The dirty images are then used to define regions for subsequent deconvolution. The dynamic range demands of spectral imaging are low enough that only a small number of \HI~detections result in severe sidelobe contamination in their respective channels. However, for visibility weighting schemes that deliver high sensitivity (e.g. for \citet{briggs1995} robustness values $r$~$>$~0), the point spread function (PSF) of MeerKAT can take on shapes that do not feature a main lobe that is well described by a 2D Gaussian \citep[see e.g.][]{jorsater1995,radcliffe2023}. In particular, the PSF can exhibit significant shoulders on the main lobe, which serve to smear out extended emission, with deconvolution necessary to provide accurate kinematic modelling and mass estimation for resolved galaxies (see also Section \ref{sec:homogenise}).

To rapidly generate deconvolution masks for the full spectral resolution deconvolution, we have developed a custom tool ({\sc pony3d}; \citet[][]{heywood2024}), which is highly parallelised, taking advantage of the fact that the images we produce remain as individual single-frequency FITS files throughout the processing, rather than monolithic cubes. The mask-making algorithm \citep[first developed by][]{tasse2021} is the same as that used by {\sc breizorro}, and the MIGHTEE continuum imaging process. Under the assumption that the image noise is Gaussian and symmetric about zero, it relates the cumulative distribution of the minima in an image to the standard deviation of the noise, and arrives at this estimate without being confused by positive true sky emission (which is less of an issue for the noise-dominated spectral line images). By drawing the distribution of minima from a sliding box (80~$\times$~80 pixels), a robust estimate of the local noise statistics can be obtained and thresholded to find genuine emission. We use {\sc pony3d} to generate a set of per-channel mask images using a local threshold of 4.8$\sigma$, which we determined to be a suitable threshold via trial and visual inspection. Although some spurious noise peaks are included in the masks due to the sheer number of pixels in the volume, at this threshold the vast majority of these can be filtered out under the assumption that any genuine spectral line emission will occupy more than one channel, which is a justifiable assumption given the high spectral resolution of the data. Single-channel events are filtered by creating three-dimensional volumes of subsets of the Boolean channel masks, and applying one iteration of binary erosion to every RA--frequency plane, followed by one iteration of binary dilation using a 3~$\times$~3 unity valued structuring element. This has the effect of removing single channel instances and leaving more extended 3D islands unaffected. The resulting islands of emission are further dilated in the spatial and spectral dimensions in order to capture any lower surface brightness extensions that a source may have. The iteration details are provided in Table \ref{tab:proc_params}. 

The resulting masks are passed as cubes to {\sc wsclean}, which is used to re-image and deconvolve the data in groups of 384 channels (or fewer for the final group). The cleaning is terminated when the peak value inside the mask drops below 1$\sigma$, where $\sigma$ is the local noise estimate as determined by {\sc wsclean}. For reasons of speed, the cleaning is restricted to a single major cycle, which is sufficient given the modest dynamic range demands of the spectral line imaging.

The above imaging procedure is repeated three times, to produce the constituent images for mosaics of all pointings at three different resolutions for each sub-band. Two of these variants simply use robustness values of $r$~=~0.0 and 0.5, while the third uses $r$~=~1.0 and applies a taper to the gridded visibilities using {\sc wsclean}'s {\sc taper-gaussian} parameter set to 40$''$. Note that this latter step results in an image cube with $\sim$1$'$ angular resolution. The imaging parameters for these three cases are listed in Table \ref{tab:proc_params}.

\subsection{Resolution homogenisation}
\label{sec:homogenise}

Each of the fifteen pointings will, for any given channel, have a range of PSF shapes, depending on the hour-angle range of the observation, RFI levels, the number of antennas in the array at that time, etc. For reasons of photometric accuracy it is highly desirable to enforce a consistent resolution across a mosaicked image. Deconvolved synthesis images have a trait that is a subtle nuisance at best, and a significant problem at worst, in that the image will be a mixture of cleaned and uncleaned emission. For the former, the model is typically restored to the map via convolution with a 2D Gaussian that is assumed to be a good fit to the main lobe of the interferometric PSF. The uncleaned (residual) emission that the convolved model is summed with remains convolved everywhere with the actual PSF. Since deconvolution is inevitably incomplete at some level, it is thus possible even in a single image for different regions of the map to have quite different `per beam' characteristics when measuring the brightness of the radio emission within.

Our method for correcting for this largely follows the one that was employed by the MIGHTEE continuum processing, except here we must apply the procedure to approximately half a million images instead of just a few. We do not enforce consistent angular resolution for any full 3D datacube as a whole, instead applying the procedure to the 15 constituent images that form each channel. The upshot of this is that the resolution advantage of the higher frequencies is retained, and any variation in resolution across a typical spectral line should be minimal (and can always be accounted for, for example via convolutional post-processing of a sub-cube). The circular 2D Gaussian beam that will be the target resolution for a given channel is decided upon by determining the maximum major axis of the fitted beams for each pointing, and adopting that value as the full-width half-maximum of the target beam. The model images for channels that have been deconvolved are then convolved with this beam. For the residual image associated with each pointing, we use the {\sc pypher} package \citep{boucaud2016} to compute a Gaussian homogenisation kernel, for which the convolution of the homogenisation kernel with the main lobe of the PSF would result in the best match to the target beam. The homogenised residual is then summed with the convolved model to produce the final per-pointing image for that channel. This process is repeated for all channels, all sub-bands, and all image-weight variations. Note that for the $r$~=~0.5 imaging the MeerKAT PSF is not well fit by a Gaussian, as the previously-mentioned PSF shoulders tend to result in a fitted beam that is significantly broader than the main lobe. For this set of images we apply a heuristic correction factor of 0.7 to the major and minor axes of the fitted beam for the above procedure (including the model convolution), which brings the fitted shape into general agreement with the shape of the main lobe. This is not an issue for the $r$~=~1.0 set of images due to the application of the additional Gaussian taper to the gridded visibilities.

\subsection{Primary beam correction and mosaicking}
\label{sec:mosaicking}

Each constituent image for a given channel is primary beam corrected by dividing the image by a model of the MeerKAT Stokes-I L-band beam generated for the specific channel frequency using the {\sc katbeam}\footnote{\scriptsize{\url{https://github.com/ska-sa/katbeam}}} package. The beam model is azimuthally averaged to remove the non-circular asymmetries that are present in the main lobe of MeerKAT's primary beam. Note that the L2 sub-band images are cut where the primary beam gain drops below 0.2, whereas the L1 sub-band is cut at a gain of 0.3. This is to include a bright \HI~source detected on the flank of the beam in the L2 sub-band, and bisected by the mosaic edge when a cut of 0.3 was applied.

A mosaic is then produced for each channel using the {\sc montage} toolkit\footnote{\scriptsize{\url{https://montage.ipac.caltech.edu/}}}, with the weighting function for each image being the square of the primary beam image. Since the noise in the primary beam corrected image scales linearly with the reciprocal of the beam model, this is equivalent to a variance weighting function for each constituent image. We produce one set of mosaicked per-channel images for both sub-bands and each of the three weighting implementations. 

\subsection{Image-domain continuum subtraction}
\label{sec:imcontsub}

The final, mosaicked spectral cubes may contain residual `continuum' emission, i.e.~features in the image that are persistent or smooth in frequency over ranges much broader than the width of a typical \HI~line. This will generally occur due to either incomplete deconvolution of the astrophysical continuum emission, or due to imperfect subtraction of a model of it due to RFI (see Sections \ref{sec:resolution} and \ref{sec:sensitivity}, and Appendices \ref{sec:flags} and \ref{sec:kurtosis}), calibration deficiencies, or other propagation effects. The cases of calibration issues or propagation effects usually amount to direction dependent effects (DDEs) which are not accounted for in the direction independent self-calibration procedure described in Section \ref{sec:selfcal}. The use of a sky model for the primary bandpass calibrator that does not include field sources and only models the central calibrator source itself has also been demonstrated to impart spectral structure into the target image data when calibrating wide field instruments such as MeerKAT and ASKAP \citep{heywood2020a}.

While the L2 band is essentially artefact-free, the L1 sub-band contains residual structures which are likely due to one or more of the effects listed above. DDEs that are caused by a combination of the antenna primary beam response combined with stochastic pointing errors can be more prevalent at lower frequencies due to the rising spectrum of a typical extragalactic radio source, and the increasing field of view of the telescope. However we also see residual features in the spectral line cube that are indicative of frequency-dependent position shifts in the apparent location of strong radio sources, which we speculate is due to the effects of ionospheric scintillation, again more prominent at lower frequencies. The characteristic positive-negative residuals are confined to small groups ($\sim$10s) of channels, and it is thus difficult to see how the primary beam (the other dominant direction dependent effect) could cause these as it is relatively smooth in frequency over such ranges.

\begin{figure*}
\centering
\includegraphics[width=0.9 \textwidth]{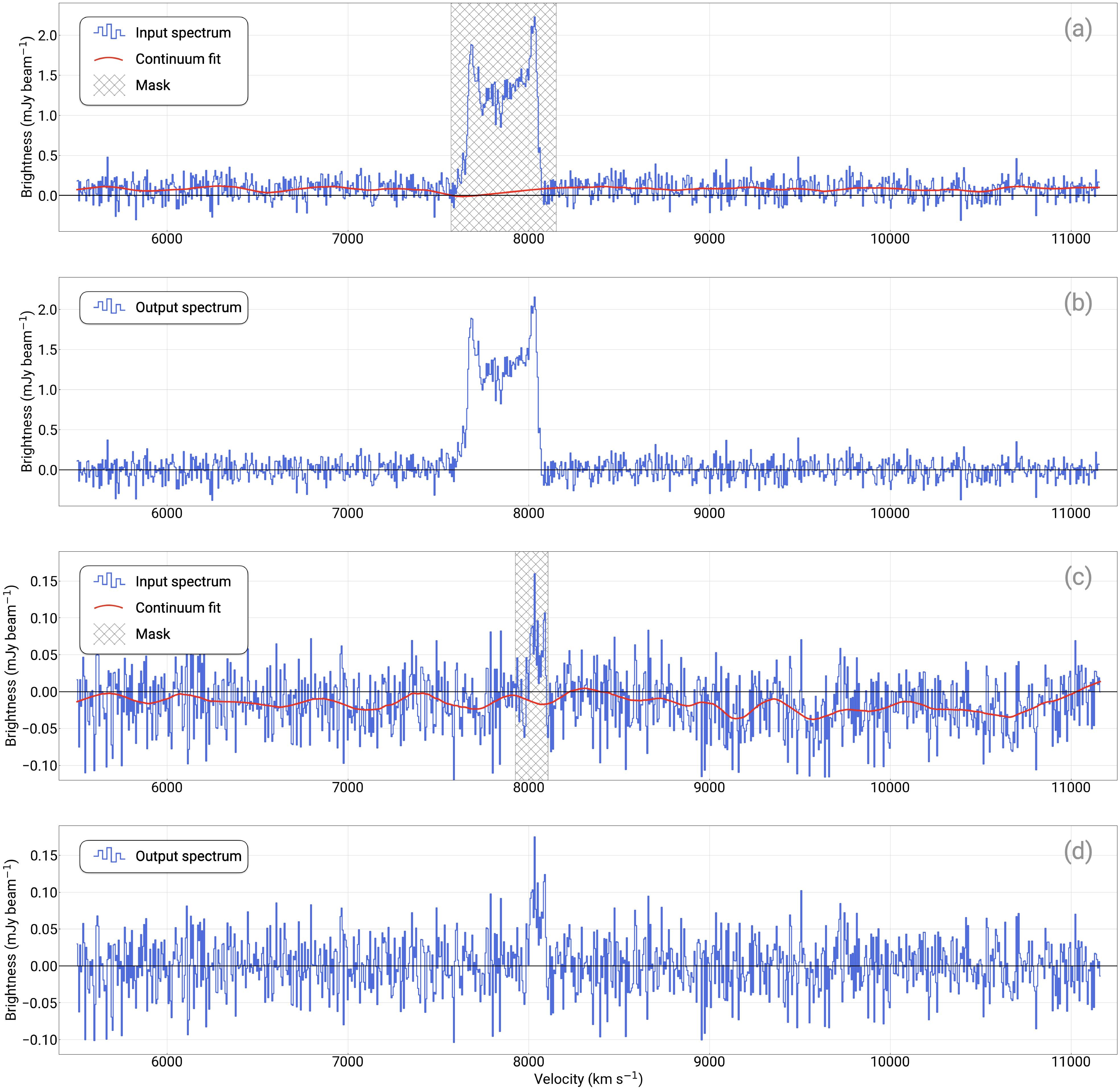}
\caption{Two examples (upper and low pairs of panels) of the image-domain continuum subtraction at work on two galaxies with differing \HI~masses. Each pair of panels shows a sightline (through a subset of channels) that intersects the \HI~disks of the two galaxies, extracted from the robust 0.5 L2 cube. The region that is automatically masked by the continuum subtraction procedure is shown by the hatched area. The upper panel for each pair (panels a and c) shows the input spectrum, with the continuum estimation shown by the continuous red line. The lower panels in each pair (b and d) show the output spectra following continuum subtraction. Note that for clarity these spectra only show 1,000 channels, approximately one eighth of the total spectral coverage of the L2 cubes. Please refer to Section \ref{sec:imcontsub} for further details.}
\label{fig:contsub_spectra}
\end{figure*}

The typical approach to dealing with such residuals is to fit smoothly-varying functions (e.g.~polynomials of low order) to the individual spectra through the cube. The method we used here involves the following steps for each spectrum:
\begin{itemize}
  \item Mask the spectrum based on five iterations of clipping on standard deviation ($\sigma$), with a threshold of $\pm$2.3$\sigma$, determined by trial and inspection of numerous spectra, both without and with \HI~lines of varying strengths.
  \item Single channel entries in the mask are removed through a single iteration of binary erosion, followed by a number of iterations of binary dilation in order to extend the mask to include any \HI~line with low level wings of emission, with the number of iterations being approximately half of the widths (in number of channels) of the filter windows used below.
  \item Masked regions are replaced in the spectrum via simple linear interpolation.
  \item The masked and interpolated spectrum is subjected to median filtering, the result of which is then passed through a Savitzky-Golay filter \citep{savitzky1964} with a polynomial order of 1.
  \item The result of the above steps is subtracted from the input spectrum, and the residual spectrum is placed back into the cube.
\end{itemize}
Sub-band specific parameters for the above procedure are summarised in Table \ref{tab:proc_params}. Figure \ref{fig:contsub_spectra} shows the effects of the image-domain continuum subtraction on two sightlines that pass through two galaxies with similar redshifts but different \HI~masses. The automatically masked region of the spectrum is indicated by the hatched areas of the input spectra shown in panels (a) and (c), with the continuum background estimation shown by the solid red lines. The resulting continuum-subtracted spectra for these two examples are shown in panels (b) and (d). The two main points to note from Figure \ref{fig:contsub_spectra} are (i) the process does not significantly bias the shape nor brightness of the \HI~line profiles; (ii) the sequential filters do not exhibit errant behaviour at the beginnings and ends of the spectra, which can be an issue for polynomial-based approaches to continuum subtraction. The removal of residual artefacts in the spectral line cubes without affecting the \HI~detections themselves greatly improves the performance of automatic source finders, and is beneficial to the L1 sub-band in particular. The effects of the image-domain continuum subtraction process on statistical studies of \HI~properties below the detection threshold \citep[e.g.][]{pan2020,chowdhury2021,pan2021,sinigaglia2022,rhee2023} remains to be seen, however the unsubtracted datacubes are a standard product of the MIGHTEE-\HI~data release, and the image-domain continuum subtraction process does not rely on retention of the visibility data, and may be subsequently refined depending on the findings of future work. We note also that the sigma clipping method that is used with the intention of excluding genuine emission or absorption features from the fit is likely to be sensitive to any source that would also be picked up by local-threshold based automatic source finder routines.

\begin{figure}
\centering
\includegraphics[width=0.8 \columnwidth]{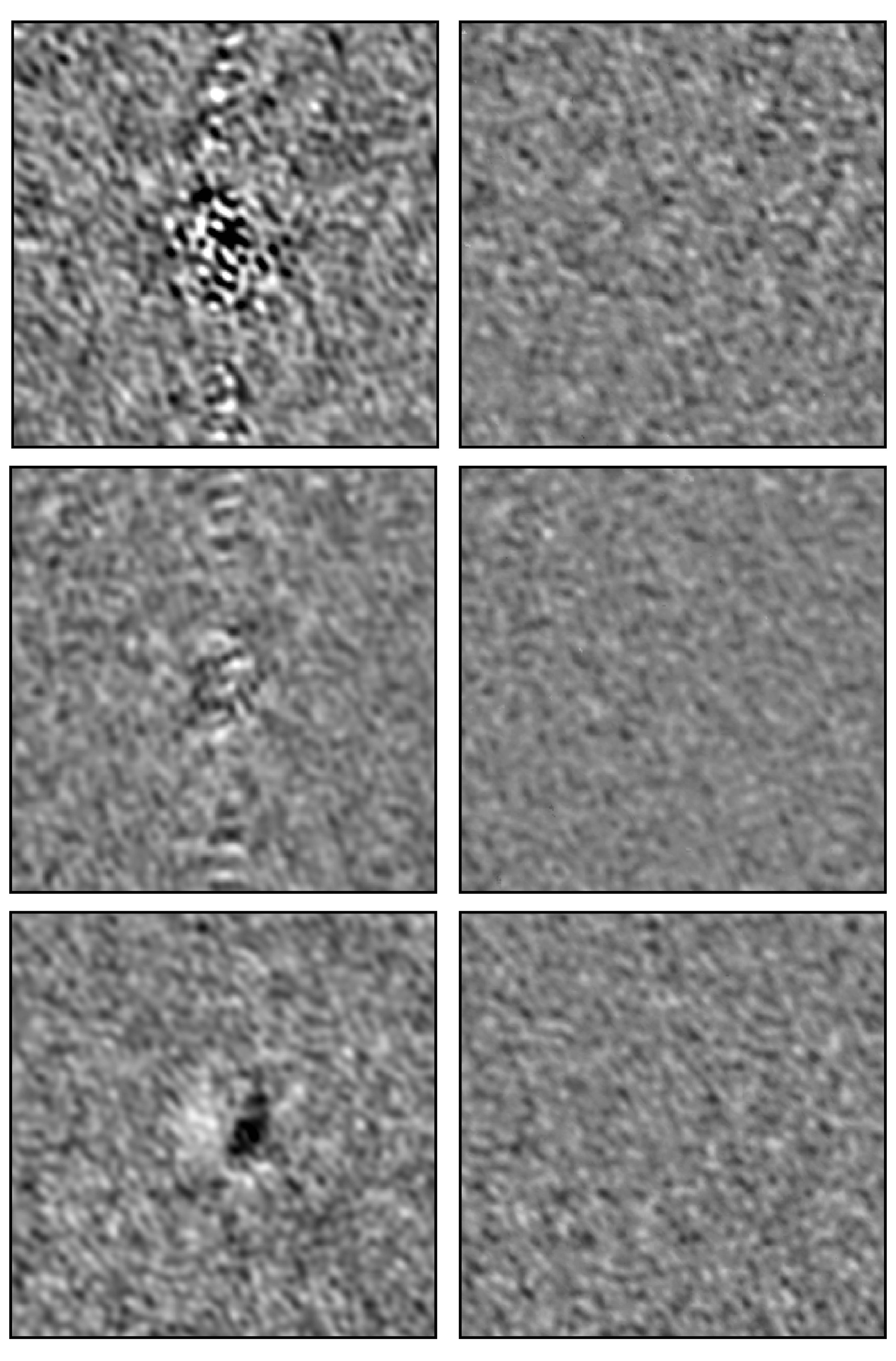}
\caption{Examples of residual imaging artefacts in the $r$~=~0.5 L1 cube from a channel at 979.3 MHz (left hand column) and the corresponding images following the image-plane continuum subtraction (right hand column). Each panel spans 9$'$, and the and the linear greyscale image runs from $-$0.5 (black) to 0.5 (white) mJy beam$^{-1}$.}
\label{fig:contsub_images}
\end{figure}

Figure \ref{fig:contsub_images} shows three examples of residual continuum artefacts, as seen in a single channel at 979.3~MHz from the L1 cube (left hand column). These artefacts are all associated with strong continuum sources at the periphery of the mosaic. The right hand column shows the same 9$'$ region following the application of the image-domain continuum subtraction algorithm described above. The first two rows are consistent with the corrupted PSF-like pattern that is associated with direction-dependent amplitude errors. The third row has a prominent positive-negative residual consistent with the subtraction of a sky model with an inaccurate position. As mentioned above, we speculate this is a manifestation of a strong phase error around the dominant source in the mosaic, introduced by ionospheric effects.

\subsection{Summary of the data products}

Each of the L1 and L2 sub-bands is imaged with three different resolution settings ($r$~=~\{0.0, 0.5, 1.0\}, with a 40$''$ taper to emphasise the core spacings applied to the latter) for a total of six sets of mosaicked spectral images. For each of these six sets we provide: (i) the dirty cubes, with no deconvolution applied; (ii) cubes with homogenised resolution within each channel; (iii) the homogenised cubes with image-domain continuum subtraction. For the L2 band (and the subset of the L1 band containing the hydroxyl maser) we also perform deconvolution, again for each of the three weighting schemes. The data release that accompanies this article therefore consists of 24 sets of spectral mosaics. The total areal coverage of the mosaics are listed for the start and end channels of each of the sub-bands in Table \ref{tab:proc_params}. 

\begin{figure*}
\centering
\includegraphics[width=0.9 \textwidth]{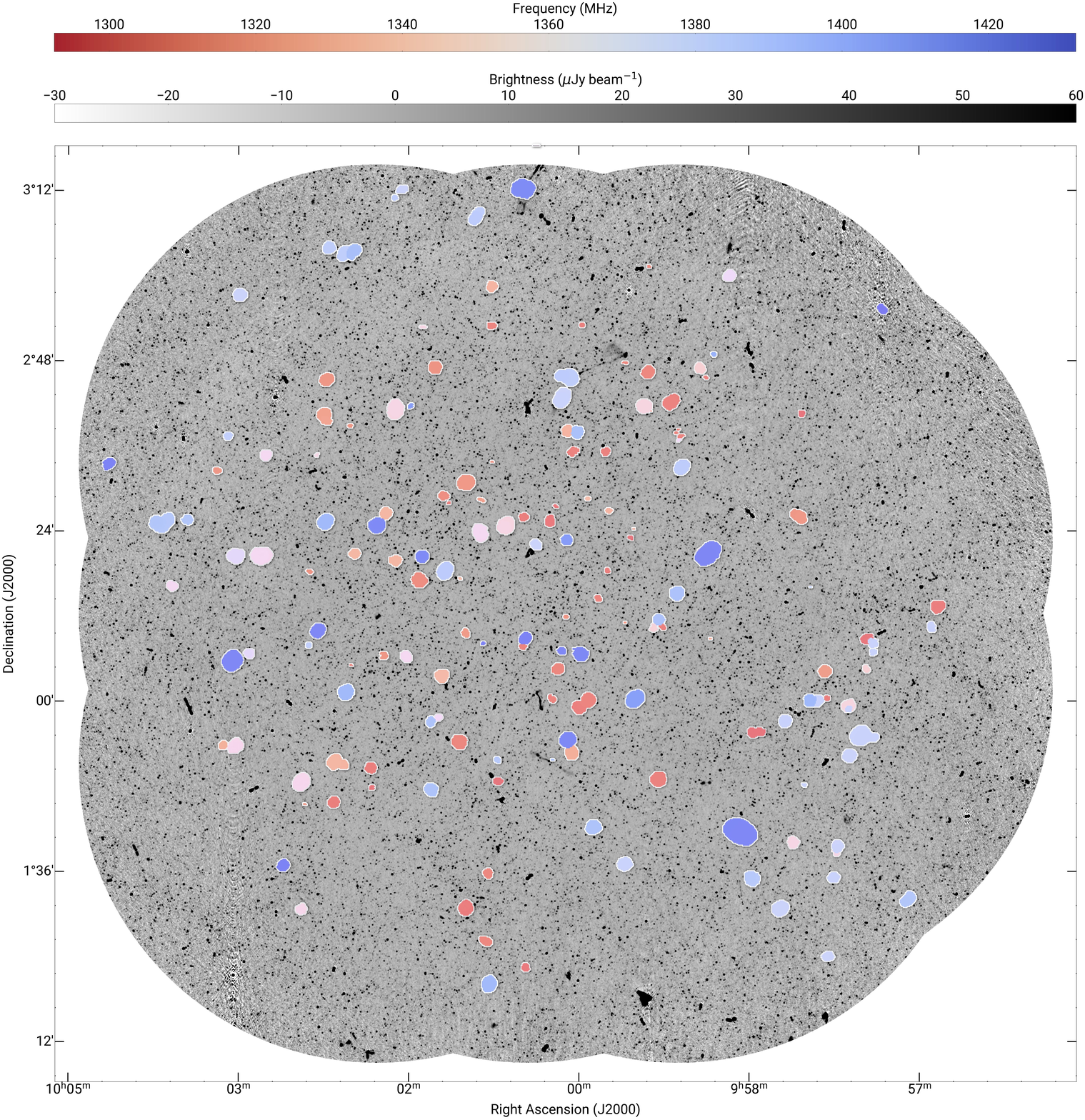}
\caption{Visualisation of a sample of \HI~detections from the deconvolved, continuum-subtracted $r$~=~1.0, tapered L2 cube. Note that the extents represent the island thresholds as identified by the {\sc pony3d} tool, and given the coarse angular resolution of the input spectral images, do not represent the true size of the \HI~disks. Detections are colour coded by the mean frequency of their contiguous 3D island of emission. The background image is the MIGHTEE continuum (Data Release 1) map (TAN projection) of the COSMOS field (Hale et al., \emph{submitted}).}
\label{fig:combo}
\end{figure*}

An image showing a sample of \HI~detections located as islands in the $r$~=~1.0 L2 data using the {\sc pony3d} tool can be seen in Figure \ref{fig:combo}. The detections are colour-coded by their frequency, and overlaid on the full total intensity MIGHTEE continuum image of the COSMOS field (Hale et al., \emph{submitted}). Like many tools that are designed to perform automatic masking or source-finding operations, {\sc pony3d} operates on the (in this case justified) assumption that regions of genuine emission can be identified as outliers against a background noise field, which is typically assumed to be Gaussian. An examination of the statistical properties of the background noise that results from our imaging procedures is provided in Appendix \ref{sec:kurtosis}. Note that a source detection catalogue will be made available with a forthcoming paper.

\section{Results and Discussion}

\subsection{Angular resolution}
\label{sec:resolution}

\begin{figure*}
\centering
\includegraphics[width=\textwidth]{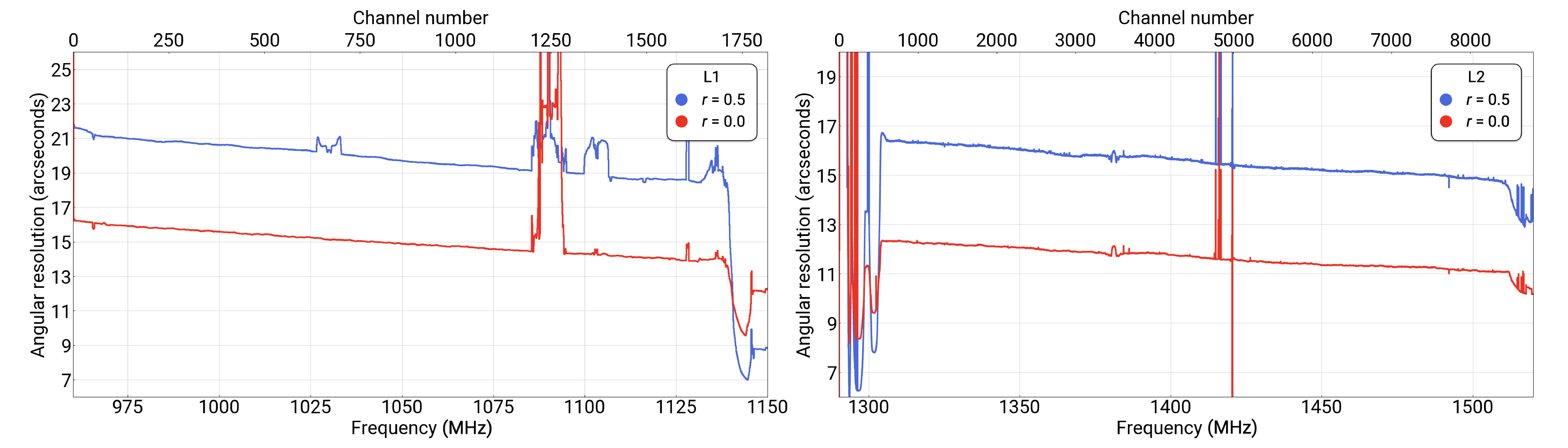}
\caption{The angular resolutions of the robust 0.0 and 0.5 data products for the L1 and L2 sub-bands as a function of frequency. Each mosaicked frequency channel is homogenised to a common circular Gaussian beam specific to that channel, as described in Section \ref{sec:homogenise}. The plots above show the full-width half-maxima of this set of beams. The per-channel frequency resolutions are 104.5~kHz and 26.1~kHz for the L1 and L2 sub-bands respectively. Note that the $\sim$arcminute angular resolution of the $r$~=~1 tapered L2 cube is omitted from this plot for reasons of clarity.}
\label{fig:resolution}
\end{figure*}

The consistent angular resolutions that are enforced for each pointing on a per-channel basis (as described in Section \ref{sec:homogenise}) are plotted on Figure \ref{fig:resolution} for the L1 (left panel) and L2 (right panel) sub-bands. The robust 0.0 and 0.5 products are plotted separately, with the resolution of the latter typically being $\sim$30 per cent lower than the former. The resolution generally varies smoothly across the band as expected, with major deviations in the L2 band occuring as the imaging enters the RFI regions at either end, as well as around the Galactic \HI~line around 1420~MHz. The true sky emission here is actually strongly affected by the autoflagging procedures, however we made no attempt to correct this given the extragalactic science goals of the MIGHTEE survey, and the locations of the target fields with respect to the Galactic plane. The other minor deviations occur at around 1380~MHz, which is the frequency used by the nuclear detonation detection system on global positioning system satellites, as well as an unidentified RFI feature at around 1490~MHz.

\subsection{Frequency dependence of sensitivity}
\label{sec:sensitivity}

\begin{figure*}
\centering
\includegraphics[width=\textwidth]{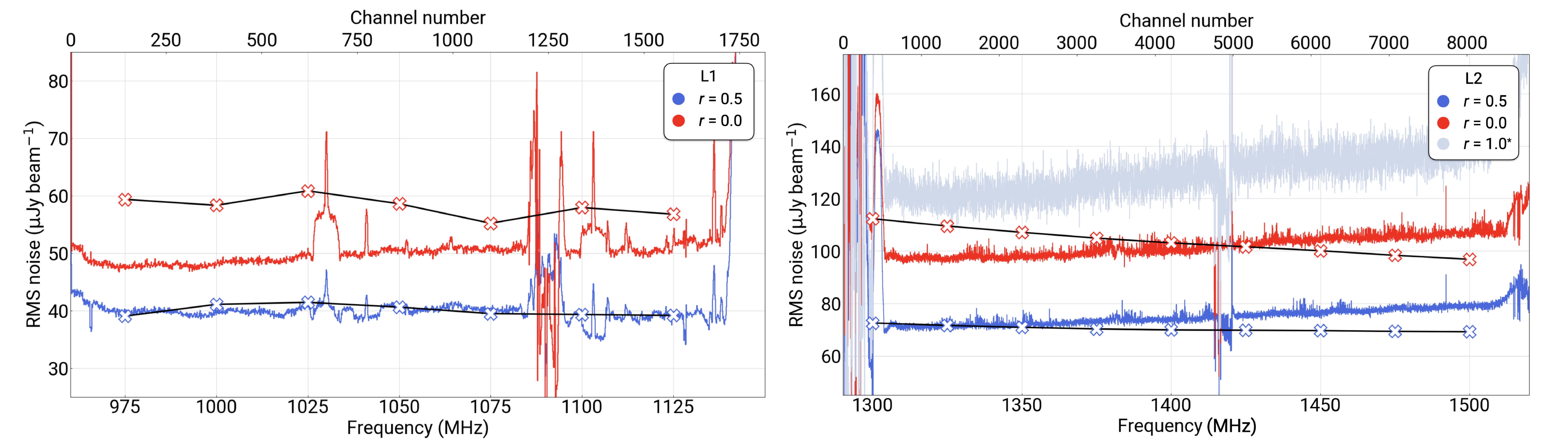}
\caption{The RMS sensitivity of the MIGHTEE~\HI~COSMOS image products as a function of frequency or channel number. The L1 (left panel) and L2 (right panel) sub-band measurements are shown from the data that does not have the image-domain continuum subtraction applied. Data produced with differing robust parameters are colour coded as per the key. The values returned by the SARAO spectral line sensitivity calculator for comparable observational parameters to ours are shown by the connected cross symbols. The per-channel frequency resolutions are 104.5~kHz and 26.1~kHz for the L1 and L2 sub-bands respectively.}
\label{fig:sensitivity}
\end{figure*}

Since we produce image products with multiple weighting / tapering schemes, the thermal noise per channel will vary. It is informative to determine how the noise in each case varies as a function of frequency, for which we present such measurements in Figure \ref{fig:sensitivity}, for the L1 and L2 sub-band cubes that have not had any image-plane continuum subtraction applied to them. Noise measurements are extracted over the region that defines the UltraVISTA area \citep{mccracken2012}, avoiding the elevated noise regions at the mosaic edges, as shown in Figure \ref{fig:schematic}. Image products with differing robust parameters are colour coded as per the inset legends. The elevated regions of noise in the L1 spectrum are principally the broad features and associated with the transmit (1030~MHz) and receive (1090~MHz) bands of the aviation secondary surveillance radar (SSR) system, and other transmissions related to distance measuring equipment (DME) for aviation (see also Section \ref{sec:chiles}, and Appendices \ref{sec:flags} and \ref{sec:kurtosis}). Aside from the edges of the L2 sub-band which encroach upon geolocation satellite transmission bands, the principal region of elevated `noise' is associated with the Galactic \HI~line around 1420~MHz. 

The connected cross symbols on Figure \ref{fig:sensitivity} show values returned by the SARAO spectral line sensitivity calculator\footnote{\scriptsize{\url{https://apps.sarao.ac.za/calculators/spectral-line}}} assuming 94~h on-source. Our achieved values are comparable to the theoretical values, although the precise assumptions that the spectral line calculator makes regarding RFI and the frequency dependence of the system temperature are not exposed to the user. A principal difference also likely arises due to our data being mosaicked. Although the mosaic is very tightly packed the effective overlap will decline with increasing frequency as the antenna primary beam pattern shrinks. Discontinuties in the predicted sensitivities are assumed to be due to the calculator's requirement for integer-valued beam sizes.

\subsection{\HI~mass sensitivity limits}

\begin{figure}
\centering
\includegraphics[width=1.0 \columnwidth]{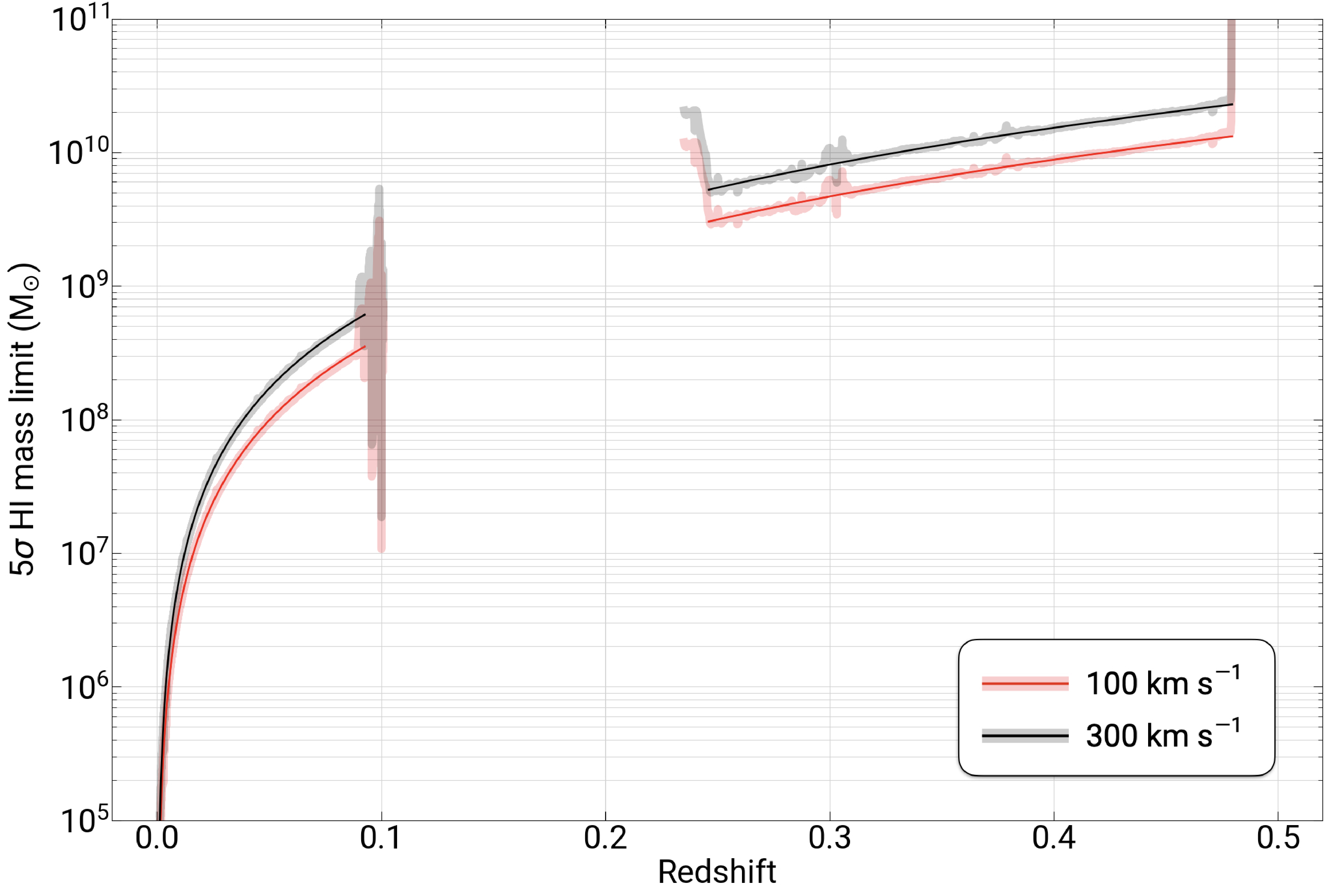}
\caption{The estimated 5$\sigma$ \HI~mass limits for the MIGHTEE-\HI~DR1 data, assuming boxcar line profiles with rest-frame velocity widths of 100~km~s$^{-1}$ (red) and 300~km~s$^{-1}$ (black). The lighter coloured traces exhibiting more variation are mass limits calculated using the measured sensitivity values (discussed in Section \ref{sec:sensitivity}), whereas the darker continuous trace uses the median 1$\sigma$ RMS channel noise values of 40 $\mathrm{\mu}$Jy beam$^{-1}$ and 74 $\mathrm{\mu}$Jy beam$^{-1}$ for L1 and L2 respectively.}
\label{fig:mass_limits}
\end{figure}

The sensitivity plots presented in Section \ref{sec:sensitivity} can be used to estimate the \HI~mass limit for the MIGHTEE-\HI~DR1 data as function of redshift. Using the methods presented by \citet{meyer2017} we determine 5$\sigma$ mass limits for the $r$~=~0.5 data for the L1 and L2 bands, assuming boxcar line profiles with widths of 100~km~s$^{-1}$ and 300~km~s$^{-1}$. The results are shown on Figure \ref{fig:mass_limits}, using both the actual per channel noise levels, as well as the median noise values (excluding the sacrificial channels at the band edges that contain entirely bad data). The latter is represented by the smooth, darker coloured line. Note that the DR1 sensitivity (and therefore \HI~mass limits) exceeds the performance of the MIGHTEE-\HI~survey as predicted by \citet{maddox2021} due to the decision to spend a higher than initially planned number of hours observing the COSMOS field.

\subsection{Positional dependence of sensitivity}
\label{sec:sens-radius}

\begin{figure}
\centering
\includegraphics[width=1.0 \columnwidth]{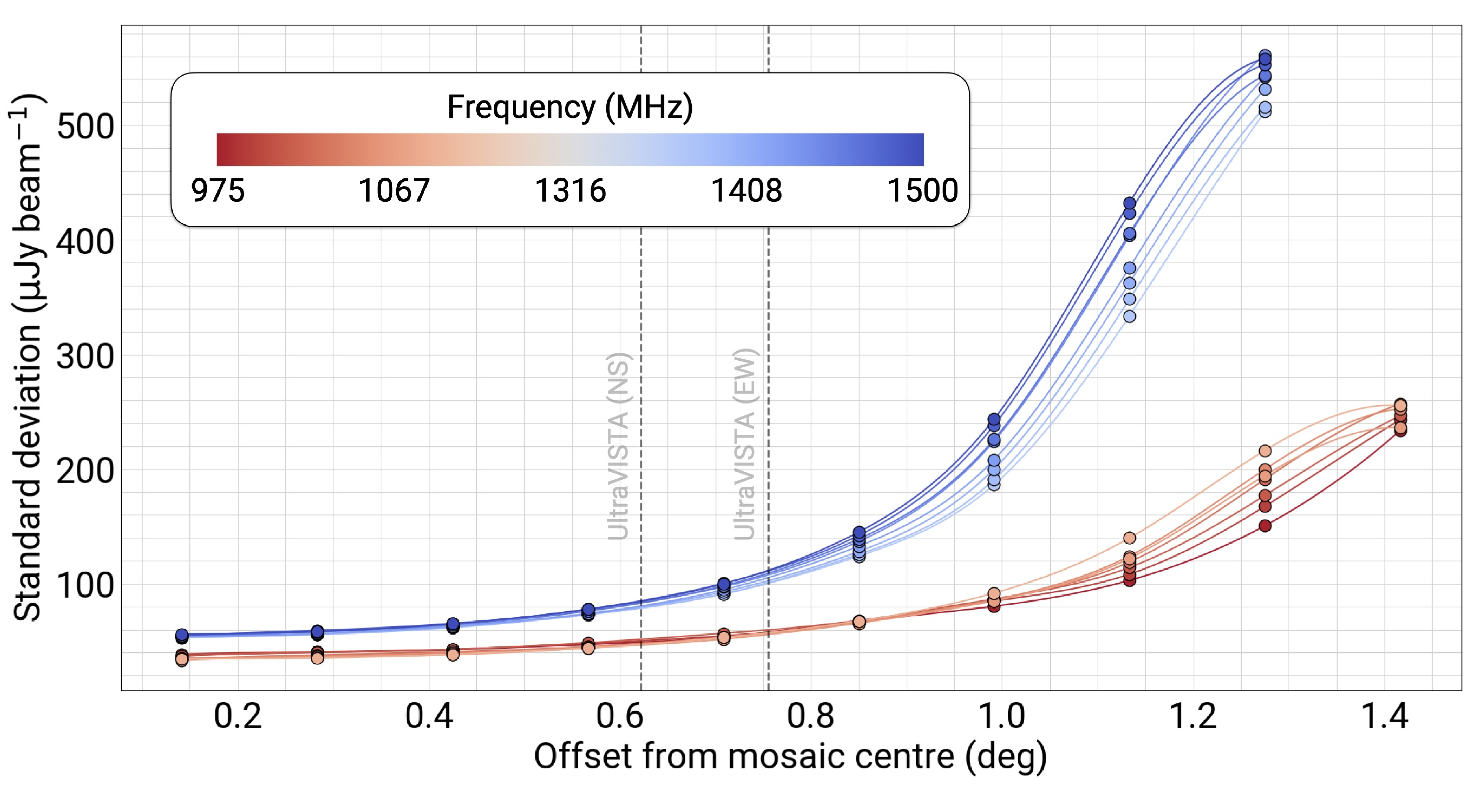}
\caption{Channel mosaic standard deviations (ordinate) as measured in concentric annuli (abscissa) for a total of 15 frequency channels across the L1 and L2 sub-bands. Spline fits are also shown on a per-frequency basis, as indicated by the inset colour bar. The north-south and east-west extents of the rectangular UltraVISTA region are shown via the vertical dashed lines.}
\label{fig:sens-radius}
\end{figure}

The mosaicking process introduces a spatial variation in sensitivity, which is dependent on the pointing layout, the antenna primary beam response pattern, and also has a frequency dependence due to that of the latter. Figure \ref{fig:sens-radius} captures these variations, by means of measurements of the standard deviation in the mosaicked channel image in concentric annuli with radial extents of 8.5$'$, centred on the image (ten annuli for L1 and nine for L2). These measurements are repeated for several frequency channels spaced equally across the band (seven for L1 and eight for L2), and the results are plotted in Figure \ref{fig:sens-radius}.

Frequency dependence is (as expected) more pronounced towards the edge of the mosaic, as indicated by the vertical spread in the points within a given annulus. The north-south and east-west extents of the deep UltraVISTA imaging are also shown on the plot. Note that detailed investigations of the \HI~mass limits and completeness of these data will be presented in a forthcoming catalogue paper.

\subsection{Examples of spatially resolved sources}

\begin{figure}
\centering
\includegraphics[width=1.0 \columnwidth]{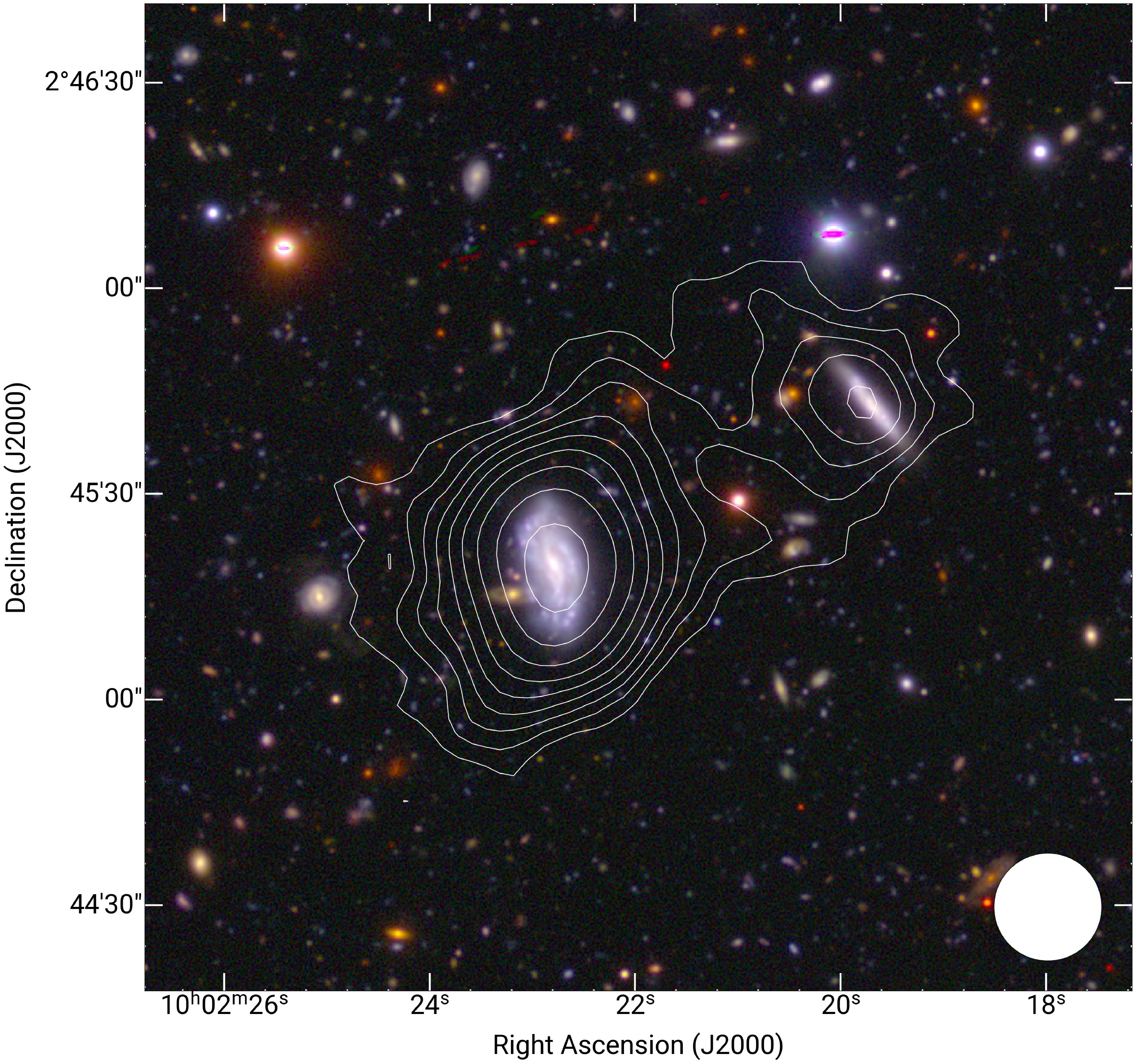}
\caption{The robust-0.5 moment-0 contour map of the \HI-detected galaxy pair J100222.8+024519 and J100219.8+024543, $z$~=~0.072266, overlaid on a 3-colour image from the $g$, $r$ and $i$ bands of the Subaru Hyper Suprime-Cam data \citep{aihara2018,aihara2019}. The contours on this figure correspond to \HI~column densities of (0.31, 0.56, 0.91, 1.4, 2.1, 3.1, 4.5, 6.35, 9.4)~$\times$~$10^{20}$~atoms~cm$^{-2}$. The circle in the lower right shows the extent of the restoring beam, which has a diameter of 15\farcs7.}
\label{fig:J100222}
\end{figure}

\begin{figure}
\centering
\includegraphics[width=1.0 \columnwidth]{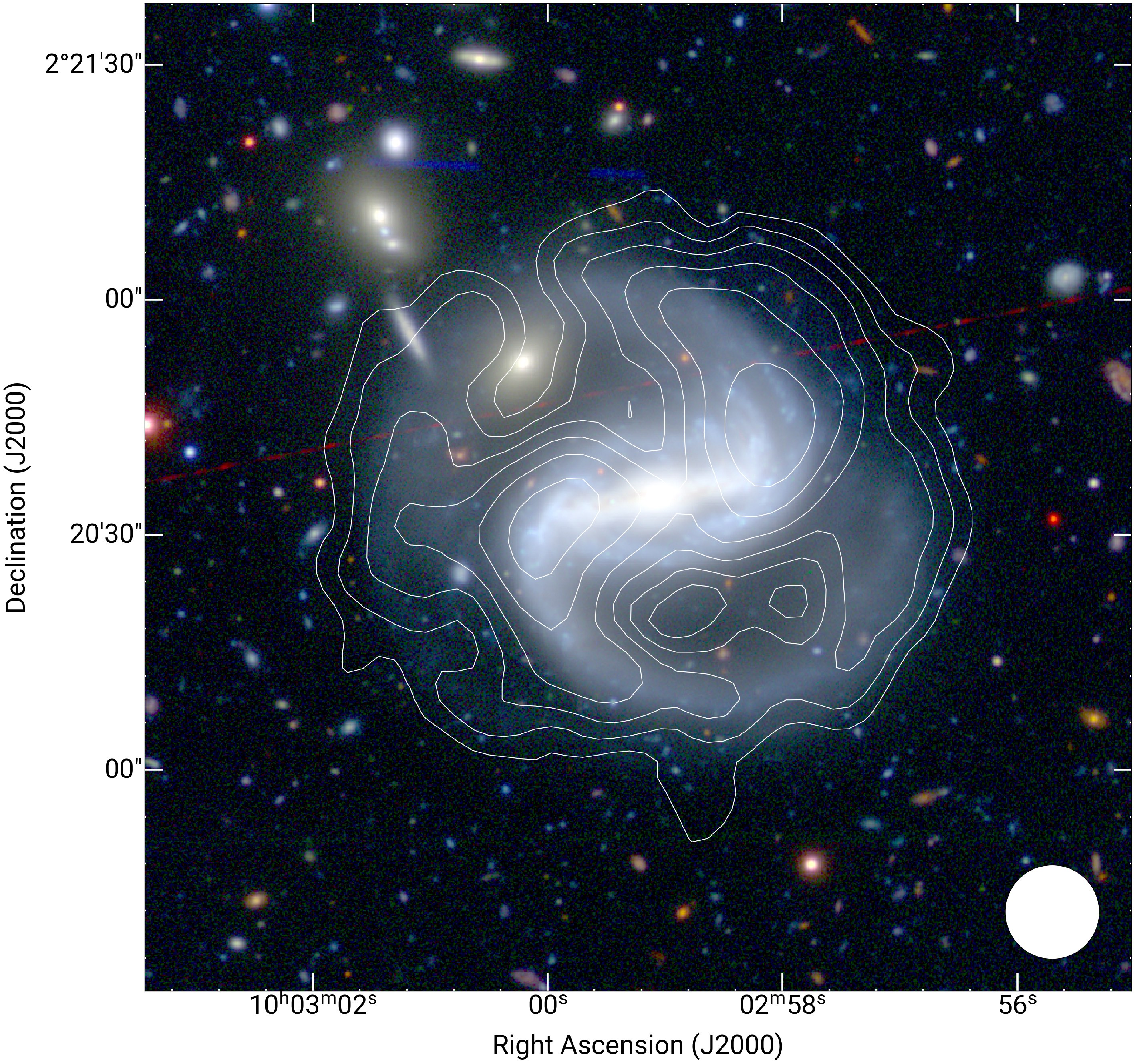}
\caption{The robust-0 moment-0 contour map of the \HI~source J100259.0+022035 at $z$~=~0.044256, coincident with a barred spiral galaxy. The background image is a 3-colour rendering of the $g$, $r$ and $i$ band photometric imaging from the Subaru Hyper Suprime-Cam data \citep{aihara2018,aihara2019}. The contours on this figure correspond to \HI~column densities of (0.56, 0.94, 1.49, 2.26, 3.35, 4.88)~$\times$~$10^{20}$~atoms~cm$^{-2}$. The restoring beam is shown in the lower right hand corner of the plot, with a diameter of 12$''$.}
\label{fig:J100259}
\end{figure}

Figures \ref{fig:J100222} and \ref{fig:J100259} show \HI~column density contour plots derived from the moment-0 maps of three spatially resolved galaxies from the L2 sub-band imaging. In both cases the background three-colour image is formed from the $g$, $r$ and $i$ band photometric imaging from the Subaru Hyper Suprime-Cam data \citep{aihara2018,aihara2019} of the COSMOS field. Figure \ref{fig:J100222} shows the \HI~emission contours associated with J100222.8+024519 and J100219.8+024543 ($z$~=~0.072266) revealing that the two spiral galaxies are interacting, and sharing a common hydrogen envelope. The \HI~emitter J100259.0+022035 ($z$~=~0.044256) shown in Figure \ref{fig:J100259} is found to be coincident with a face-on barred spiral galaxy. The resolved \HI~disk is seen to follow the spiral arms, with a build up of hydrogen at the ends of the barred potential, whereupon the interstellar medium likely predominantly transitions to the molecular phase \citep[e.g.][]{combes2019}. 

These two images serve as an example of the benefits of imaging the data with multiple weighting schemes. The data in Figure \ref{fig:J100222} are from the robust 0.5 imaging. Although the angular resolution of the robust 0.5 set of images is $\sim$30 per cent lower than the robust 0.0 imaging (see Figure \ref{fig:resolution}), the additional sensitivity to lower surface brightness features means the common envelope is readily detected. Both the lower \HI~mass galaxy and the bridging feature are not readily visible in the robust 0.0 images, which are both noisier (see Section \ref{sec:sensitivity}) and less sensitive to diffuse structures. However the \HI~contours in Figure \ref{fig:J100259} make use of the robust 0.0 data, where the higher resolution imaging is beneficial for resolving the spiral structure in the \HI~emission, as well as the two peaks of emission representing the accumulation of \HI~at the ends of the bar.

\subsection{Derived \HI~mass comparisons}

\begin{figure}
\centering
\includegraphics[width=1.0 \columnwidth]{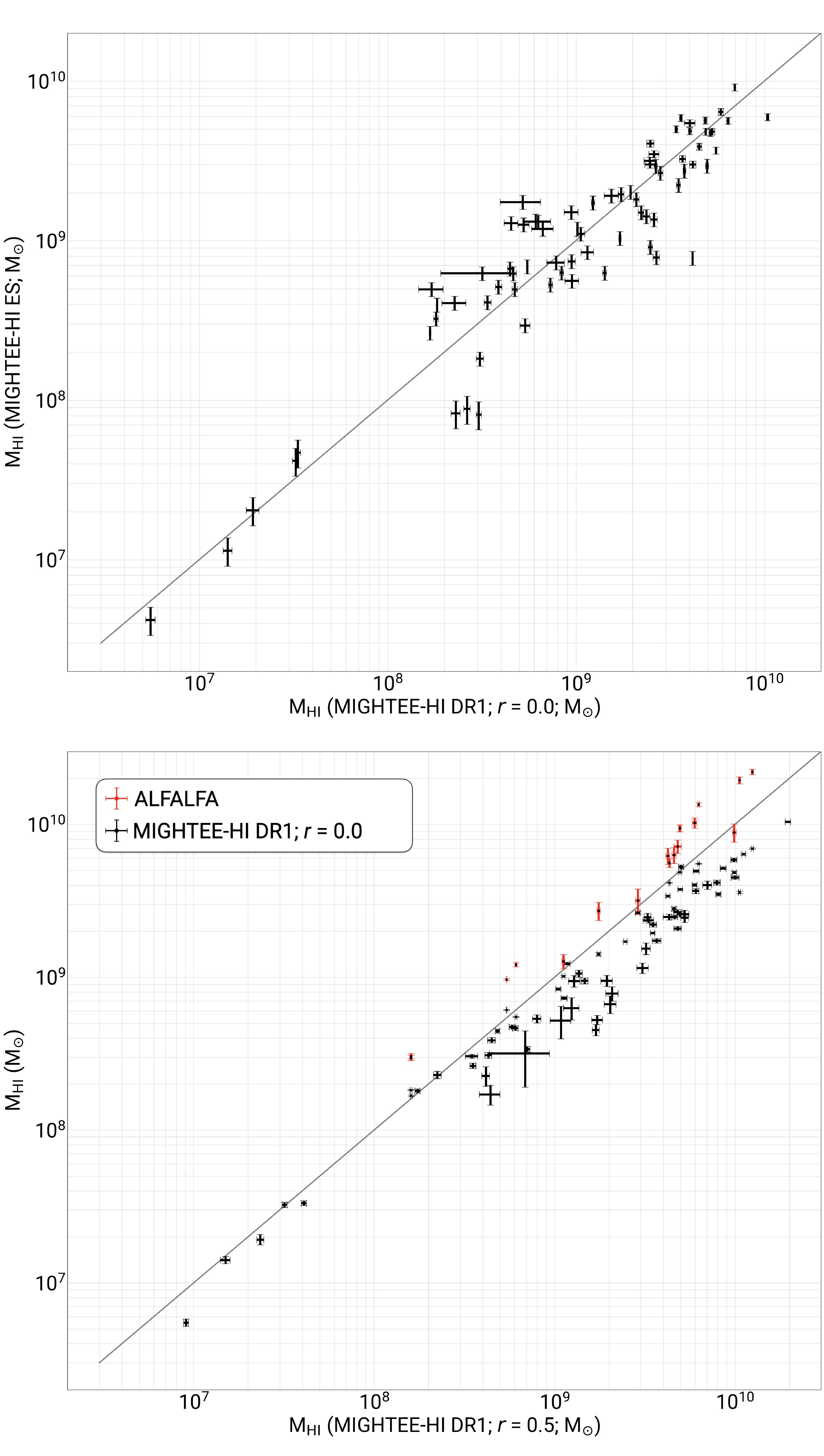}
\caption{The \HI~mass measurements for 17 galaxies from ALFALFA (red markers) and 75 galaxies from the MIGHTEE-\HI~Early Science data (black markers). The upper panel shows the MIGHTEE-\HI~Early Science data against the corresponding measurements for the same galaxies detected in the $r$~=~0.0 data presented in this article. The lower panel shows the $r$~=~0.0 mass measurements (black markers) and ALFALFA mass measurements (red markers) against those made from the $r$~=~0.5 images.}
\label{fig:mass-mass}
\end{figure}

In Figure \ref{fig:mass-mass} we compare total \HI~mass measurements derived from some of the DR1 detections to matched values from existing measurements. Specifically, we compare 75 detections to the measurements from the MIGHTEE-\HI~Early Science data \citep[upper panel;][]{ponomareva2023}, which has already been used for numerous studies as summarised in Section \ref{sec:intro}, and 17 detections to the corresponding measurements from the Arecibo Legacy Fast ALFA (ALFALFA) survey \citep[lower panel;][]{haynes2018}. We also compare the \HI~mass measurements derived from the $r$~=~0.0 and $r$~=~0.5 DR1 images. Total integrated \HI~line fluxes are measured for each galaxy in the DR1 data, and converted to \HI~mass measurements following the methods of \citet{meyer2017}. The error bars in the case of both sets of MeerKAT measurements are estimated in a consistent way, by projecting the three-dimensional mask describing the extent of a given \HI~galaxy to 50 emission-free regions surrounding it in the cube. The total integrated line flux in each of these regions is measured, and the uncertainty in the galaxy's integrated line flux density is then taken to be the standard deviation of the measurements in these 50 blank regions \citep[for further details see][]{ponomareva2021}. These flux error estimates are then propagated into the mass calculation. The uncertainties in the ALFALFA measurements are taken directly from the extragalactic \HI~catalogue of that survey.

\begin{figure}
\centering
\includegraphics[width=1.0 \columnwidth]{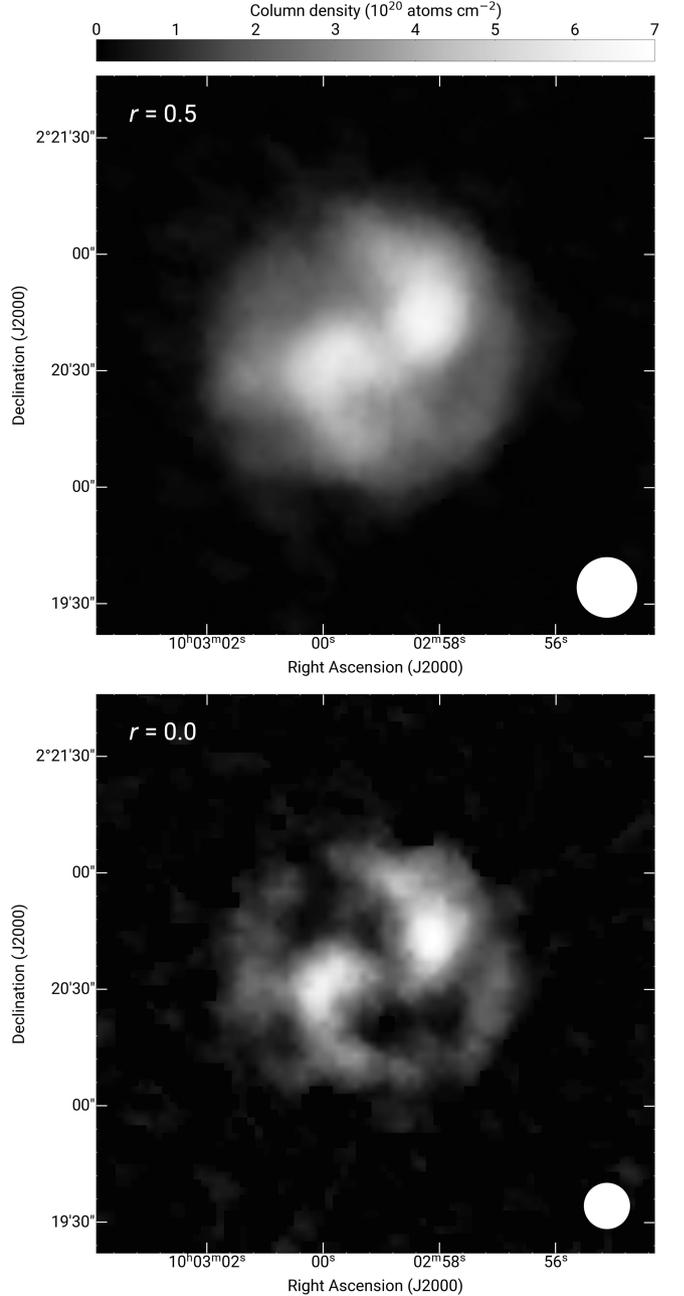}
\caption{The moment-0 \HI~column density maps for J100259.0+022035, for the $r$~=~0.5 imaging (upper panel) and the $r$~=~0.0 imaging (lower panel). The restoring beam sizes are shown by the white circles in the lower right hand corner of each plot. Significantly more \HI~is recovered by the $r$~=~0.5 imaging (log$_{10}$~$M_{\mathrm{HI}}$~=~9.69 M$_{\odot}$ and 9.99 M$_{\odot}$for $r$~=~0.0 and 0.5 respectively), as the angular resolution is traded for increased sensitivity (in terms of both image noise and sensitivity to extended emission).}
\label{fig:robusts}
\end{figure}

The cross-matched measurements exhibit some scatter from the 1:1 line, however the differing bulk behaviour of the two external samples when compared to our new results is readily explainable (ignoring the possibility of calibration differences, which should be minimal between the two sets of MeerKAT observations). The MIGHTEE-\HI~Early Science data has beam sizes comparable to those of the DR1 $r$~=~0.0 data, and as the upper panel of Figure \ref{fig:mass-mass} shows the two sets of data are in good agreement. Differences are likely to arise in the total \HI~mass measurements since the DR1 data are marginally deeper but have 8$\times$ higher spectral resolution than the Early Science data, and no deconvolution was performed for the latter. The lack of deconvolution will likely have a significant effect since the majority of the comparison sample are spatially resolved. The $r$~=~0.5 data systematically recovers more \HI~than the $r$~=~0.0 data (0.2 dex on average), as can be seen by the black markers in the lower panel of Figure \ref{fig:mass-mass}. This is to be expected as higher sensitivity is exhanged for lower angular resolution as the Briggs' robustness parameter becomes more positive, and more weight is given to the shorter baselines in the array. Figure \ref{fig:robusts} emphasises this point. The upper panel shows the $r$~=~0.5 moment-0 \HI~column density map (log$_{10}$~$M_{\mathrm{HI}}$~=~9.99 M$_{\odot}$) for J100259.0+022035, with the corresponding $r$~=~0.0 (log$_{10}$~$M_{\mathrm{HI}}$~=~9.69 M$_{\odot}$) image shown in the lower panel.

In the case of the ALFALFA measurements, a single dish telescope is more likely to detect the total \HI~line flux, hence the excess seen in the corresponding points, particularly when using a signal-to-noise limited cut to define the emission in the interferometric images. Note also that a comparison between the DR1 $r$~=~0.5 data and measurements from observations of the COSMOS field with the Five-hundred-meter Aperture Spherical radio Telescope (FAST) also show no systematic deviations between the measured masses in a cross-matched sample over $>$3 orders of magnitude in $M_{\mathrm{HI}}$ \citep{pan2024}. In the case of the FAST comparison presented in that work, the aperture used for the MeerKAT flux density extraction corresponds to the size of the FAST beam.

\subsection{Non-detection of the CHILES $z$~=~0.3764 emission line}
\label{sec:chiles}

\begin{figure*}
\centering
\includegraphics[width=0.9 \textwidth]{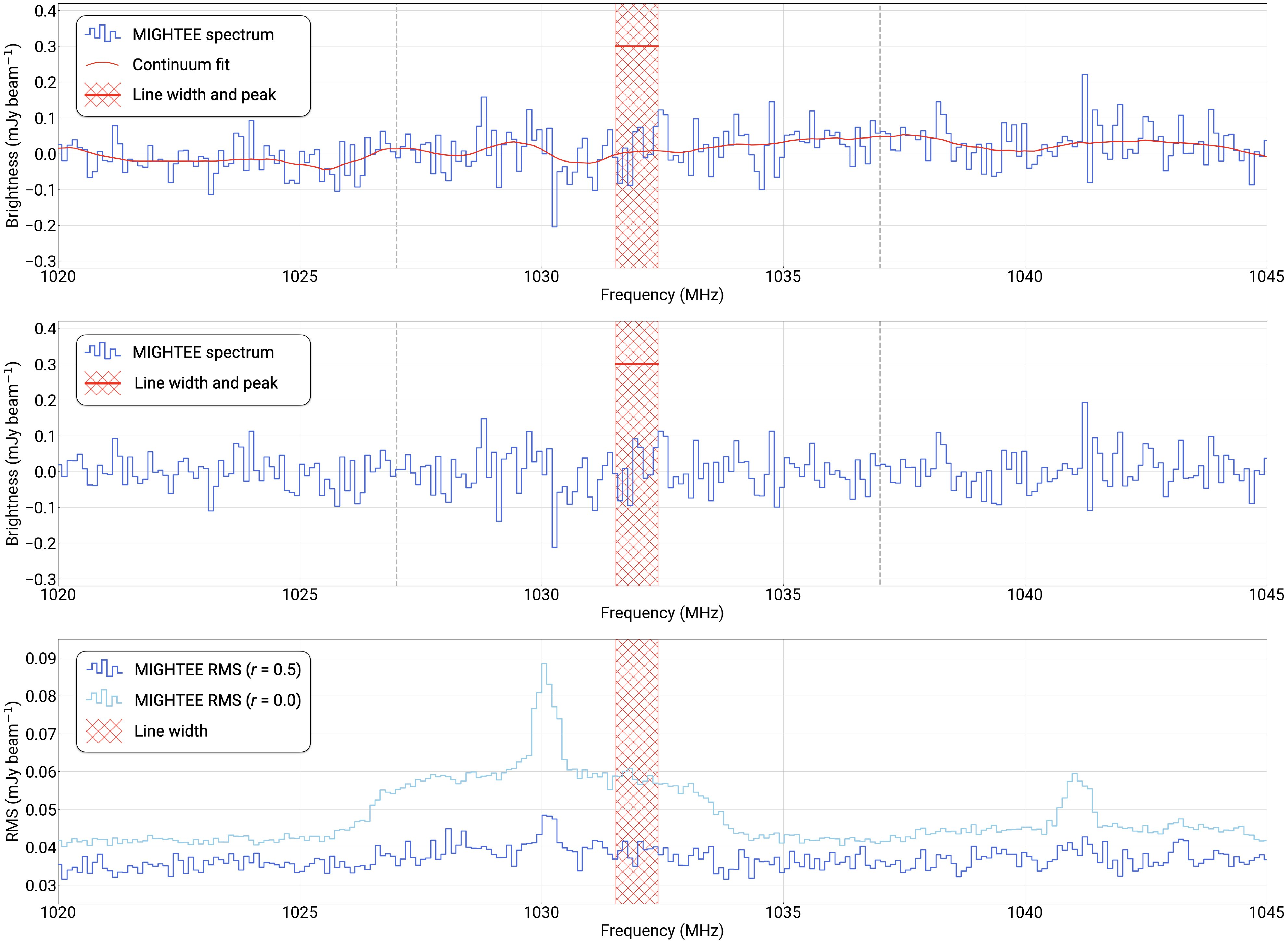}
\caption{The upper panel shows a spectrum from the L1 ($r$~=~0.5) sub-band, extracted at the position of the $z$~=~0.3764 \HI~detection from the CHILES survey reported by \citet{fernandez2016}. The width (red hatched region) and peak (red horizontal bar) of the reported detection are also shown. The continuous red trace is the image-domain continuum subtraction's estimation of the spectral baseline. The central panel shows the spectrum that results following subtraction of the estimate. The lower panel shows the RMS of the MIGHTEE data, measured over a circular aperture of 9$'$ diameter centred on the same position for the $r$~=~0.5 data (dark blue) as well as the $r$~=~0.0 data (pale blue) in order to emphasise the presence of the 1030 MHz secondary surveillance radar transmissions used for aviation.} 
\label{fig:chiles}
\end{figure*}

Our data are deep enough to be able to study the $z$~=~0.3764 \HI~detection from the COSMOS \HI~Large Extragalactic Survey \citep[CHILES;][]{fernandez2013}. As the highest redshift \HI~emitter reported to date, this object is of interest, as it represents an opportunity to obtain a direct \HI~mass measurement in a galaxy at cosmological look-back time of over 4~Gyr. It could potentially be detected in the MIGHTEE-\HI~data with  improved surface brightness sensitivity (and lower angular resolution) over the existing B-configuration observations from the Karl G.~Jansky Very Large Array (VLA), possibly refining such a measurement.

The detection as presented by \citet{fernandez2016} has a central frequency of 1031.967 MHz with a width of 0.875 MHz (246 km~s$^{-1}$ rest-frame) at position J100054.83+023126.2, and a peak of 0.3 mJy beam$^{-1}$ \citep[see also][]{fernandez2023}. The reported line emission is partially spatially coincident with a spiral galaxy, but exhibits a fainter extended tail of \HI~emission that extends 10$''$ in a southern direction. The redshift of the galaxy was confirmed via optical and millimetre-wave emission to also be coincident with the \HI~peak in the CHILES spectrum.

To search for corresponding emission in the MIGHTEE data we extract a spectrum at the aforementioned position, coincident with the spiral galaxy. The spectrum is extracted from the $r$~=~0.5 cube, which at this frequency has a resolution of 20$''$ (see Figure \ref{fig:resolution}). We thus expect the vast majority of the \HI~emission from this system to be spatially unresolved. Note that we do not expect astrometric frame mismatches to be an issue here. All tests of the MIGHTEE data to date have revealed offsets that are a small fraction of a radio map pixel in size, both from comparison to radio data from the VLA \citep{heywood2022a} and optical data \citep{whittam2024} (see also Figures \ref{fig:J100222} and \ref{fig:J100259}).

We also extract a corresponding spectrum from the cube with the image-domain continuum subtraction. These two spectra are shown in the upper two panels of Figure \ref{fig:chiles}. The red hatched area and bar marker also show the peak and width of the reported line detection. As can be seen, no significant emission is found in the spectrum, and the image-domain continuum subtraction procedure does not affect this result.

The lower panel of Figure \ref{fig:chiles} shows the RMS of the MIGHTEE data as measured in a 9$'$ diameter region centred on the galaxy position for both the $r$~=0.5 data (upon which the analysis in this section is based), as well as the $r$~=~0.0 data for which the longer baselines are emphasised relative to the more numerous shorter ones. \citet{fernandez2016} note the proximity of the line to a region of RFI, which we identify here as the 1030~MHz SSR interrogation frequency (see also Section \ref{sec:sensitivity}). We also note that this transmission has a pedestal-like spectral shape that is very prominent in the $r$~=~0.0 data, that affects several MHz of the L-band spectrum, including the range of the CHILES detection. There is no clear signature of this feature in either the corresponding $r$~=~0.5 spectrum, or visually in the images themselves. 

\begin{figure}
\centering
\includegraphics[width=1.0 \columnwidth]{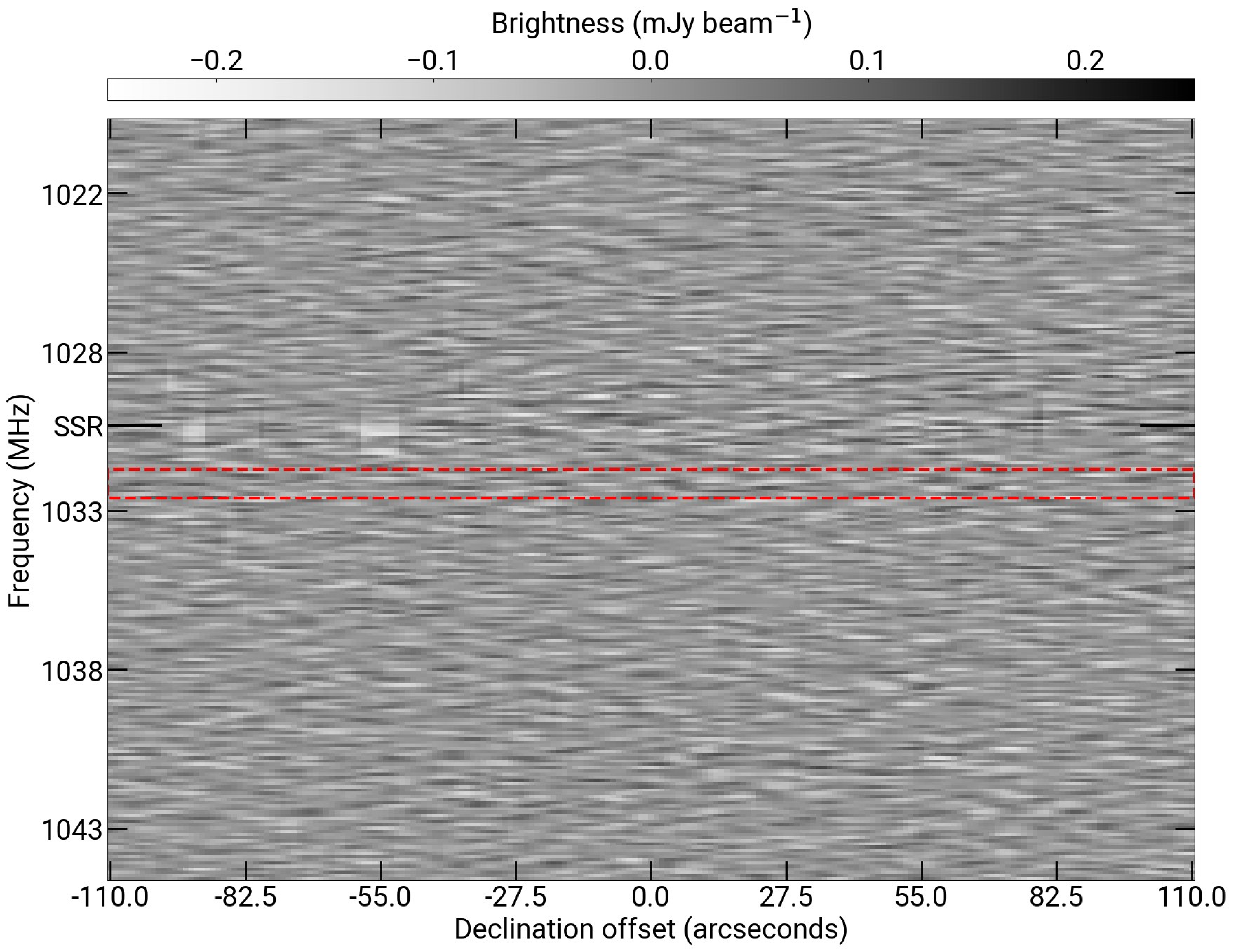}
\caption{A frequency / declination slice through the $r$~=~0.5 MIGHTEE-\HI~data, centred at RA~=~10\hhh00\mmm54\sss.83, Dec~=~+02\ddd31\mmm26\farcs2. The frequency range is the same as that covered in Figure \ref{fig:chiles}, and the declination slice covers $\pm$110$''$, or $\pm$5 synthesised beam widths. The 1030~MHz SSR transmission is marked with the large tick mark, and the dashed lines show the reported width of the $z$~=~0.37 line emitter.}
\label{fig:pv}
\end{figure}

Additionally, a plot of frequency against declination for the $r$~=0.5 data is shown in Figure \ref{fig:pv}. The frequency range covers the same region of the band as Figure \ref{fig:chiles}, with declination offsets of $\pm$100$''$ ($\pm$5 synthesised beam widths), centred on RA~=~10\hhh00\mmm54\sss.83, Dec~=~+02\ddd31\mmm26\farcs2. A declination slice was chosen due to the orientation of the \HI~tail seen in the VLA data. Some low level artefacts associated with the 1030~MHz SSR transmissions are visible, but no sigificant emission is seen at the position of the reported \HI~line.

Examining the amplitudes of a cube containting Fast Fourier Transforms of the channel images (equivalent to the gridded visibility amplitudes as a function of frequency) shows that the RFI feature does not preferentially affect any particular region of MeerKAT's $u,v$ plane, however given that the source is unresolved in the 20$''$ cube, a disproportionate loss of the shorter spacings should not affect its detectability. A detailed summary of the flagging statistics in our data is provided in Appendix \ref{sec:flags}.

We evaluate (non-)detection probabilities given our data by taking the line width and the \HI~mass with its associated uncertainty reported by \citet{fernandez2016}. We generate 100,000 boxcar profiles with the appropriate width, determine the spectrum that would result from a \HI~mass drawn from a normal distribution with a mean of 2.9~$\times$~10$^{10}$~M$_{\odot}$ and a standard deviation of 1.0~$\times$~10$^{10}$~M$_{\odot}$, and perturb each frequency channel with Gaussian noise consistent with our measured per-channel noise values in the vicinity of the position. We then compare the resulting set of synthetic spectra to the corresponding measurements from the MIGHTEE-\HI~cube, and count the instances where the spectra meet certain criteria. The probabilities that a galaxy with these parameters would be detected in the MIGHTEE data with frequency-integrated detection significances of 3$\sigma$ and 5$\sigma$ are 98.3 per cent and 94.8 per cent respectively. We emphasise that these figures are conservative, since the VLA B-configuration data purports to spatially resolve the \HI~emission, thus a lower angular resolution observation from a telescope with superior low surface brightness sensitivity should detect such emission at higher significance (see the lower panel of Figure \ref{fig:mass-mass} for a demonstration of such an effect). We place a 3$\sigma$ upper limit on the \HI~mass for this system of 8.1~$\times$~10$^{9}$~M$_{\odot}$, assuming a boxcar line profile with a width of 246 km~s$^{-1}$.

\section{Conclusions}

We have used 94.2~h of (on-source) MeerKAT L-band time to produce the first public data release (DR1) for the spectral line component of the MIGHTEE survey. The products are formed via independent, consistent processing of the visibility databases from 15 individual pointing centres covering the COSMOS field, which are then linearly mosaicked and further processed in the image-domain. We synthesise image products with various visibility weighting schemes in order to trade sensitivity for angular resolution. As we have demonstrated in this article, the 12$''$ resolution cubes will allow resolved studies of morphology and kinematics out to higher redshifts than has been previously possible, over $>$4 deg$^{2}$ of sky coverage. We provide deconvolved products for the L2 sub-band, and the newly-discovered $z$~=~0.7012 hydroxyl megamaser discovered in the L1 sub-band \citep{jarvis2024}. Cubes with image-domain continuum subtraction are also provided. The continuum subtraction method is shown to not affect our direct \HI~detections, but it may be refined in future depending on the outcome of pending statistical studies of the spectral line emission close to or below the thermal noise limit. These will be executed by exploiting with the wealth of multiwavelength data available in the target fields of the MIGHTEE survey.

The achieved noise levels of the DR1 data are comparable to the noise values predicted by SARAO's line sensitivity calculator for similar observational parameters. The public DR1 data products offer significant improvements over the initial, internal MIGHTEE-\HI~Early Science data release, which has itself been successfully used for numerous studies of the cold gas properties of galaxies within a subset of the final MIGHTEE survey volume. We attempted to perform a follow-up study of the highest redshift detection of \HI~in an unlensed galaxy \citep[$z$~=~0.376;][]{fernandez2016}, however we find no evidence of spectral line emission at the reported position. We posit that the original detection is likely to be spurious, possibly due to its proximity to the 1030~MHz aviation secondary surveillance radar transmission, the effects of which are readily seen in various properties of our data.

The data processing strategy outlined in Section \ref{sec:dataproc} now forms the basis for future processing of the remainder of the high spectral resolution MIGHTEE L-band observations that are already in-hand, namely 36, 56, and 10 tracks (of either 4 or 8 h duration) in the XMM-LSS, CDFS, and ELAIS-S1 fields respectively. Prior experience with the continuum processing \citep[][Hale et al., \emph{submitted}]{heywood2022a} suggests that these fields are likely to have higher dynamic range demands for spectral line imaging than the COSMOS field, due to the presence of strong radio sources in the field of view that may not be adequately subtracted using direction-independent self-calibration methods. At L-band frequencies direction dependent effects are predominantly primary-beam and pointing error related, however the lower frequency sub-band imaging of the COSMOS field contains residual structures that are suggestive of ionospheric effects. Our processing workflow is amenable to the addition of direction-dependent processing schemes for the subtraction of problematic bright sources directly from the visibilities, as will likely be necessary for future MIGHTEE-\HI~data releases.

\section*{Acknowledgements}

We are very grateful to Jacqueline van Gorkom for reviewing this paper and substantially improving it in the process. The MeerKAT telescope is operated by the South African Radio Astronomy Observatory, which is a facility of the National Research Foundation, an agency of the Department of Science and Innovation. We acknowledge the use of the ilifu cloud computing facility – www.ilifu.ac.za, a partnership between the University of Cape Town, the University of the Western Cape, Stellenbosch University, Sol Plaatje University and the Cape Peninsula University of Technology. The Ilifu facility is supported by contributions from the Inter-University Institute for Data Intensive Astronomy (IDIA – a partnership between the University of Cape Town, the University of Pretoria and the University of the Western Cape, the Computational Biology division at UCT and the Data Intensive Research Initiative of South Africa). The authors acknowledge the Centre for High Performance Computing (CHPC), South Africa, for providing computational resources to this research project. We thank Masechaba Sydil Kupa, Chris Schollar, Fernando Camilo and Jeremy Smith for facilitating the hosting of the data products. This research made use of {\sc astropy},\footnote{\scriptsize\url{http://www.astropy.org}} a community-developed core Python package for Astronomy \citep{astropy:2013, astropy:2018}. This work has made use of the Cube Analysis and Rendering Tool for Astronomy \citep[{\sc carta};][]{comrie2021}. This research made use of {\sc montage}, which is funded by the National Science Foundation under Grant Number ACI-1440620, and was previously funded by the National Aeronautics and Space Administration's Earth Science Technology Office, Computation Technologies Project, under Cooperative Agreement Number NCC5-626 between NASA and the California Institute of Technology. This research has made use of NASA's Astrophysics Data System. We acknowledge the indispensability of {\sc matplotlib} \citep{hunter2007}, {\sc numpy} \citep{harris2020}, and {\sc scipy} \citep{virtanen2020}. IH, MJJ, and AAP acknowledge support of the Science and Technology Facilities Council (STFC) grants [ST/S000488/1] and [ST/W000903/1]. IH, MJJ and HP acknowledge support from a UKRI Frontiers Research Grant [EP/X026639/1], which was selected by the European Research Council. IH acknowledges support from the South African Radio Astronomy Observatory which is a facility of the National Research Foundation (NRF), an agency of the Department of Science and Innovation. IH acknowledges support from Breakthrough Listen. Breakthrough Listen is managed by the Breakthrough Initiatives, sponsored by the Breakthrough Prize Foundation. IH thanks Nadeem Oozeer and D.J. Pisano for useful discussions. MJJ acknowledges generous support from the Hintze Family Charitable Foundation through the Oxford Hintze Centre for Astrophysical Surveys. MG is supported by the Australian Government through the Australian Research Council’s Discovery Projects funding scheme (DP210102103). SHAR is supported by the South African Research Chairs Initiative of the Department of Science and Technology and the National Research Foundation. IR acknowledges support from the STFC grant [ST/S00033X/1]. MGS acknowledges support from the South African Radio Astronomy Observatory and National Research Foundation (Grant No. 84156). MV acknowledges financial support from the Inter-University Institute for Data Intensive Astronomy (IDIA), a partnership of the University of Cape Town, the University of Pretoria and the University of the Western Cape, and from the South African Department of Science and Innovation's National Research Foundation under the ISARP RADIOMAP Joint Research Scheme (DSI-NRF Grant Number 150551) and the CPRR HIPPO Project (DSI-NRF Grant Number SRUG22031677).

\section*{Data Availability}

The raw visibility data are available from the SARAO archive by searching for the capture block IDs listed in Table \ref{tab:obs}. The image products described in this article are available at \url{https://doi.org/10.48479/jkc0-g916}.





\appendix

\section{Flagging levels}
\label{sec:flags}

\begin{figure*}
\centering
\includegraphics[width=\textwidth]{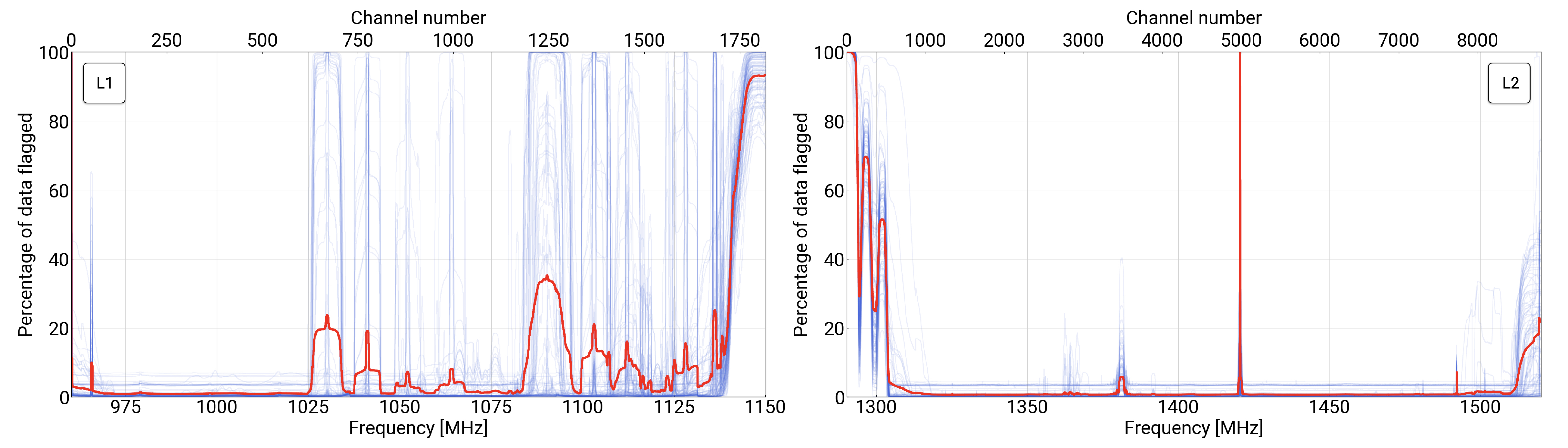}
\caption{The percentage of data flagged as a function of frequency channel. These measurements are made for each of the seven scans that make up a single pointing, for a total of 105 independent scans. The results for each scan are shown by the faint blue histograms. The flag percentages for the COSMOS observations as a whole are shown by the red histogram. This process is repeated for both the L1 (left panel) and L2 (right panel) sub-bands.}
\label{fig:flagchan}
\end{figure*}

\begin{figure*}
\centering
\includegraphics[width=\textwidth]{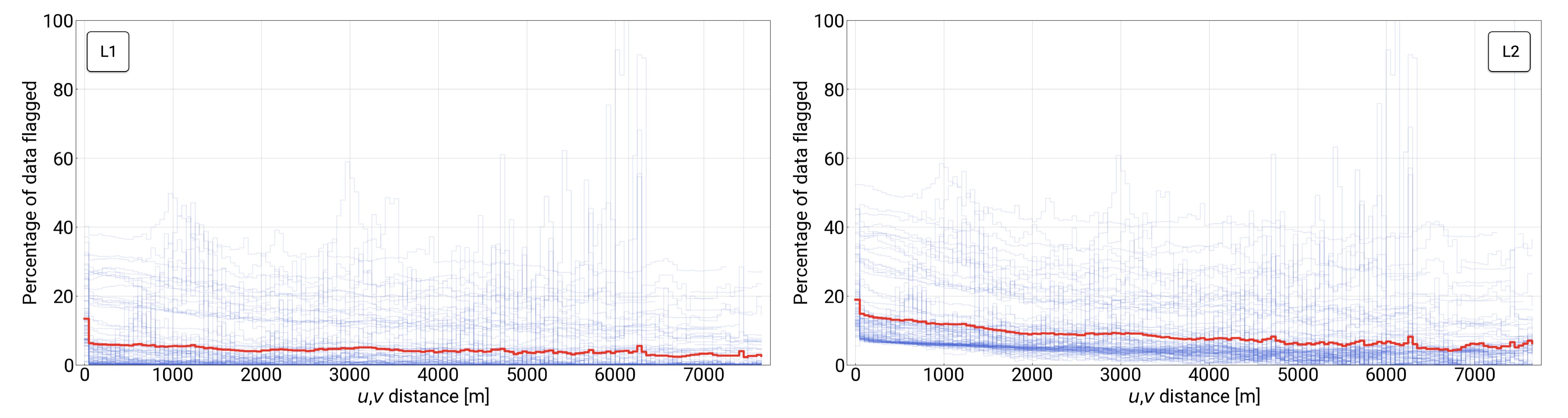}
\caption{The flagging percentages as a function of $u$,$v$ distance for the L1 (left panel) and L2 (right panel) sub-bands. The blue and red histograms represent the individual scans and the overall percentages for the COSMOS observations as a whole, as per Figure \ref{fig:flagchan}. Note that for these plots the channels at the L1 and L2 band edges where the data encroach on the persitent satellite RFI bands are excluded.}
\label{fig:flaguvdist}
\end{figure*}

The percentage of data flagged as a function of frequency and $u$,$v$ distance (i.e.~baseline length) are shown in Figures \ref{fig:flagchan} and \ref{fig:flaguvdist}. The fifteen MeerKAT pointings that were observed at full spectral resolution make up the MIGHTEE-\HI~COSMOS field, with each of these observations containing seven contiguous scans of the target field, each scan being approximately 55 minutes in duration. The flagging statistics are determined on a per-scan basis, for a total of 105 unique scans.

The spectrum for the L2 band has very low flag percentages, with the exception of the emission from the 1380~MHz nuclear detonation detection system, the Galactic \HI~line, the unidentified feature at 1490~MHz, and the sacrificial channels at the edges of the L2 band where the data extends into the persistent bands of the geolocation satellites. Similar edge features are present at the ends of the L1 band. The L1 band is much more prone to RFI, largely due to transmissions associated with aviation, namely the SSR and DME systems. Several scans exhibit high levels of flagging in the aviation-related bands 1025~MHz, serving to significantly elevate the mean flagging percentages for the observations as a whole. Further investigation reveals that the vast majority of these are associated with the COSMOS\_1, COSMOS\_2, COSMOS\_3, and COSMOS\_4 pointings, which were all observed within 1 month of each other. We can only speculate as to the origin of this, possibly high levels of aircraft activity at that time. Note that a 35 per cent data loss (corresponding to the mean flag percentage associated with the strongest SSR feature at 1090~MHz) results in only a 20 per cent reduction of the channel map sensitivity.

\begin{figure}
\centering
\includegraphics[width=1.0 \columnwidth]{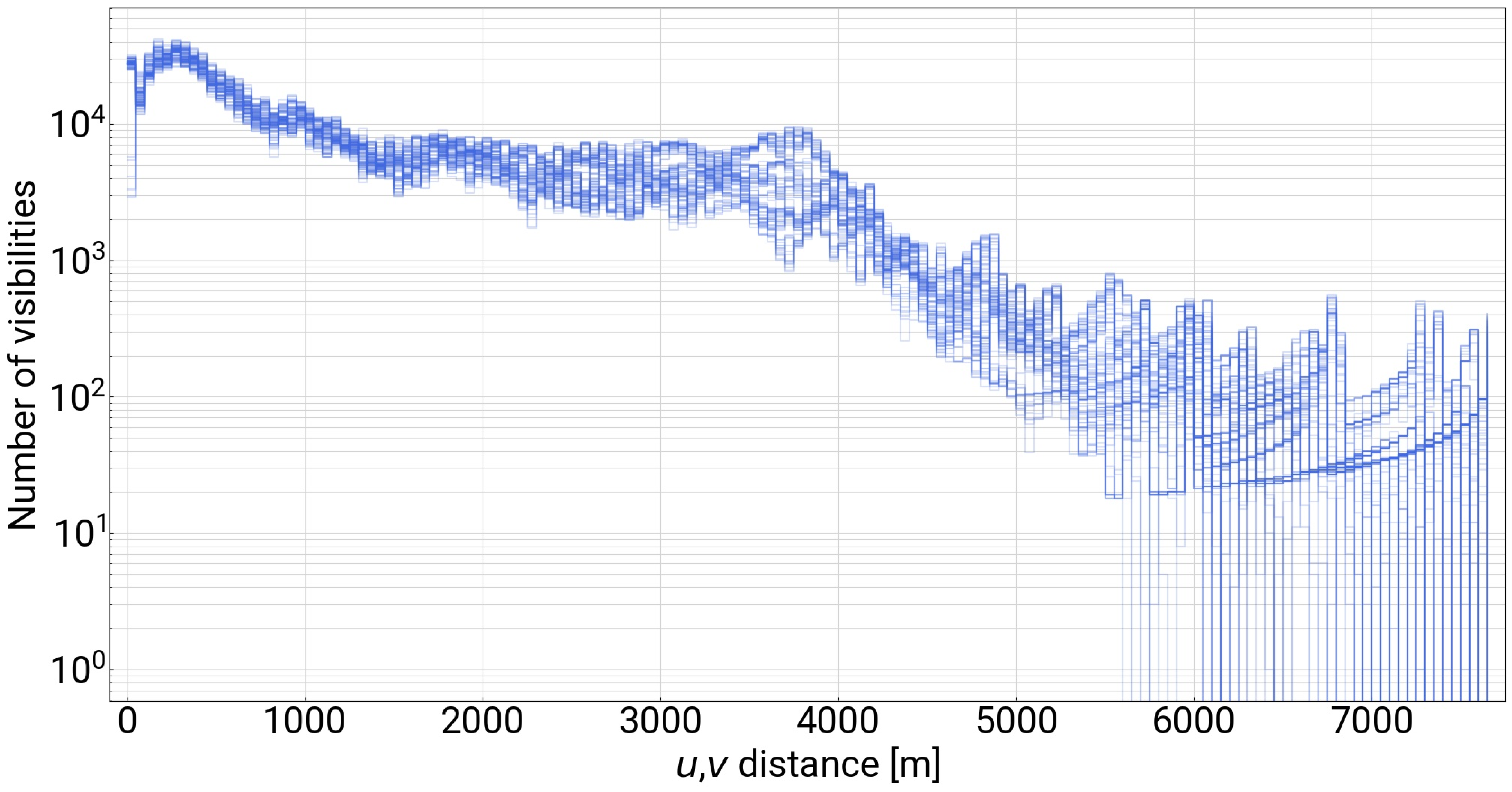}
\caption{The number of visibility measurements in 154~$\times$~50~m $u$,$v$ bins for the 105 individual scans that make up the COSMOS spectral line mosaics.}
\label{fig:viscounts}
\end{figure}

Figure \ref{fig:flaguvdist} shows the percentages of data lost to flagging, measured in 154 $u$,$v$ distance bins of 50~m each. Note that the flagging levels are determined as a percentage of the number of visibilities in each specific bin. MeerKAT's dense core layout means that in real terms the data loss is much more significant for shorter baselines. To illustrate this, the number of visibilities as a function of baseline length for these 154 bins is shown in Figure \ref{fig:viscounts}, where it can be seen that the innermost spacings in the array contain $\sim$3 orders of magnitude more visibilities than the outermost.

The average percentage losses remain relatively flat as a function of baseline length for the L1 band, with a mild increase towards the shorter spacings in L2; the destructive effects of RFI thus do not have a strong dependence on baseline length, and the inner region of the $u$,$v$ plane is only affected by virtue of there being more visibility measurements in that region, and not due to the shorter baselines being intrinsically more likely to detect RFI. This picture is reinforced by the behaviour of the flagging histograms for the individual scans. The four pointings that are blighted by high levels of aviation-related RFI again bring up the mean for the set, however there is no major short-spacing excess in the flagging levels. We note the tendency of a small number of scans to suffer total and inexplicable data loss at around 6000~m in both the L1 and L2 bands, although (noting the distribution of visibility counts in Figure \ref{fig:viscounts}) one of the outermost antennas being entirely flagged could be a probable explanation. Note that the flagging statistics as a function of baseline length are determined with the band edge features excluded.

\section{Image noise properties}
\label{sec:kurtosis}

Feature extraction algorithms used for e.g.~source finding and automasking for deconvolution generally make the assumption that such features can be identified as statistically significant outliers against a backdrop of Gaussian noise. Deviations of this background from Gaussianity can thus affect the performance and completeness of such techniques.

For a single pixel feed array such as MeerKAT, only the visibility measurements have independent, uncorrelated noise characteristics. Once the data are inverted into the image domain the pixels in the image are spatially correlated, as the image (including the noise) is convolved everywhere with the PSF (or in the case of a deconvolved emission, a mixture of the PSF and a Gaussian restoring beam). However, in the absence of calibration errors, incomplete deconvolution, or source confusion, the pixel value distribution of the background noise in a synthesised radio image can be closely represented by a Gaussian distribution, particularly for observations where the PSF does not have highly structured sidelobes.

\begin{figure*}
\centering
\includegraphics[width=\textwidth]{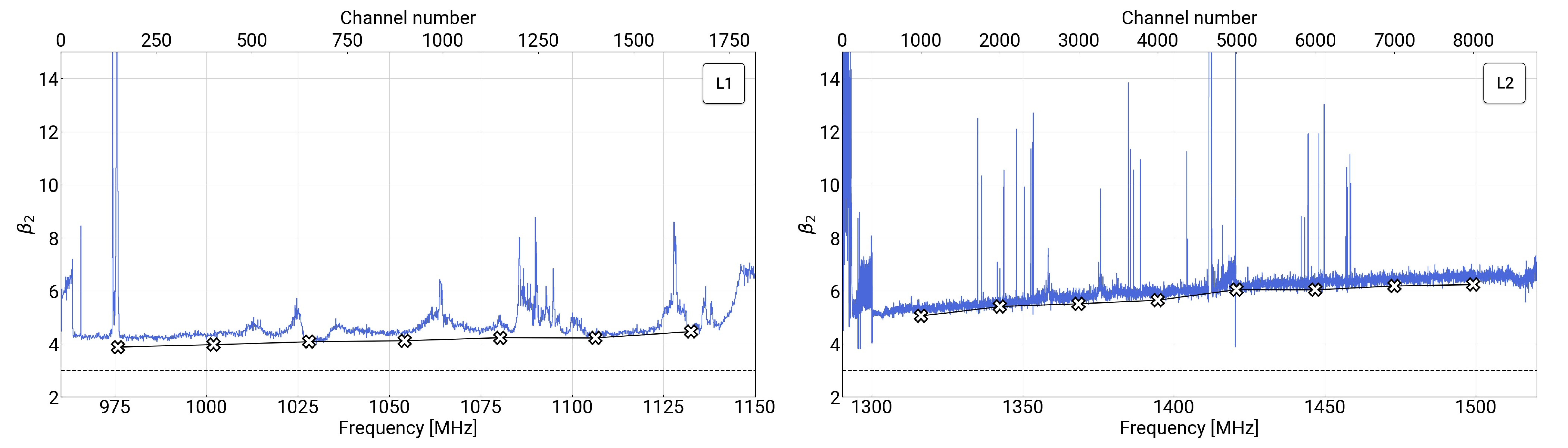}
\caption{The Pearson kurtosis ($\beta_2$) values determined over the UltraVISTA region (see Figure \ref{fig:schematic}) as a function of frequency for the L1 and L2 sub-bands. The causes of the elevated regions are discussed in Section \ref{sec:kurtosis}. The cross markers and solid line represent the ideal case for our data, whereby the imaging and mosaicking process has been repeated for visibility data that is pure Gaussian noise. The dashed horizontal line shows the $\beta_2$~=~3 value that represents a purely Gaussian distribution of image pixels.}
\label{fig:kurtosis_full}
\end{figure*}

Here we examine the noise properties of our mosaicked image products by means of the Pearson kurtosis value ($\beta_2$), for which a purely Gaussian distribution of values will have $\beta_2$~=~3. Figure \ref{fig:kurtosis_full} shows the $\beta_2$ measured within the UltraVISTA region of the mosaics as a function of frequency for the L1 and L2 sub-bands, with the dashed line showing the $\beta_2$~=~3 of pure Gaussian noise. With reference to the L1 plot, the high $\beta_2$ spike at 975 MHz is the $z$~=~0.7012 OH maser \citep{jarvis2024}, with other elevated values corresponding to aviation-related transmissions, namely SSR and DME. In the L2 plot, contiguous sets of channels (which still appear narrow at the resolution of the plot) with higher kurtosis values occur where there are strong \HI~detections. Isolated spikes in the kurtosis also occur when a single channel is missing from one of the constituent pointings, which serves to bring the edge of the mosaic closer to the optical box area, thereby raising the noise level.

As mentioned above, a synthesised radio image has pixels that are correlated on the scale of the synthesised beam (PSF). To determine the effects this has on the kurtosis values we replace the visibilities of the 15 constituent pointings with pure Gaussian noise, then repeat the imaging and mosaicking process. The kurtosis values of this ``empty'' sky are then determined as above. The results of this process at seven frequencies in L1 and eight in L2 are shown by the crosses and lines on Figure \ref{fig:kurtosis_full}, demonstrating that for channels that are not affected by residual RFI or genuine astrophysical emission, the noise statistics for our data are very close to the ideal case.

\begin{figure*}
\centering
\includegraphics[width=\textwidth]{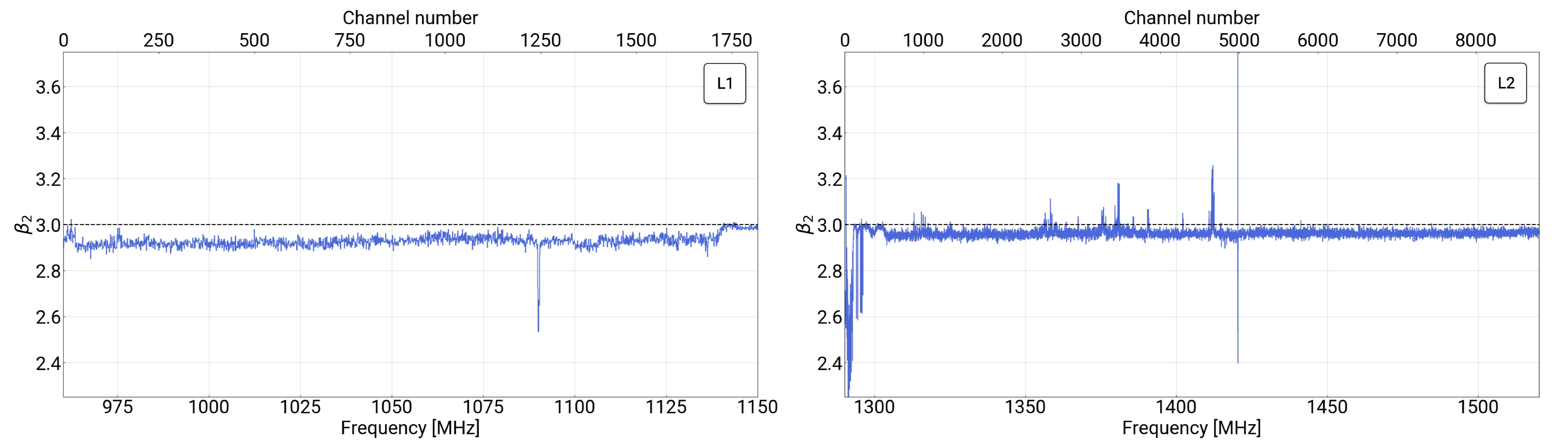}
\caption{The mean Pearson kurtosis ($\beta_2$) values for the L1 and L2 sub-bands as a function of frequency, as measured from 1,000 regions covering 80~$\times$~80 pixels within the mosaicked image that forms each channel in our data release. The fact that in all cases the $\beta_2$ values are close to the ideal Gaussian value of 3 suggests that non-Gaussianity of the pixel statistics is unlikely to affect any source finding or mask making algorithms that make use of sliding boxes (i.e. the vast majority) to determine the positional dependency of the background noise levels.}
\label{fig:kurtosis_box}
\end{figure*}

A second test is also performed, which is probably more relevant in terms of how any non-Gaussianity of the noise may affect the performance of source finding and automasking techniques. Such software (including the {\sc pony3d} tool described in Section \ref{sec:imaging}) generally employs a sliding box method to estimate the positional dependence of the background noise. We thus measure the kurtosis in 1,000 regions of 80~$\times$~80 pixels and take the mean. This is repeated for each channel, and the results are shown in Figure \ref{fig:kurtosis_box} for the L1 and L2 bands. In both cases the $\beta_2$ values are close to the pure Gaussian noise values.

Note that in all cases above, the $\beta_2$ values are measured from the $r$~=~0.5 data, which will represent the worst case scenario in terms of our data products due to its higher PSF sidelobe levels.

\section{Image products}

Table \ref{tab:data} summarises the image data products that form the MIGHTEE-\HI~DR1. The data release consists of 89 three dimensional FITS cubes. The L1 data cubes cover the full channel range, however the L2 sub-band is provided as nine individual cubes with 55 overlapping duplicate channels. Frequency ranges for each cube are provided in Table \ref{tab:data}, along with the maximum redshift for \HI~and the 1665~MHz OH maser line. The $z$~=~0.7012 megamaser reported by \citet{jarvis2024} is provided in separate 199 channel sub-cubes. The various labels that appear in the file names describe properties of a particular cube as follows:

\begin{itemize}

    \item{{\bf dirty:} This means that the (up to) fifteen constituent images that form a single channel mosaic have their own individual point spread function, and have not been modified in any way.}

    \item{{\bf conv:} The constituent images have been homogenised according to the procedure described in Section \ref{sec:homogenise}.}

    \item{{\bf clean:} Spectral line emission has been located and deconvolved following the steps outlined in Section \ref{sec:imaging}.}

    \item{{\bf contsub}: The image-domain continuum subtraction method described in Section \ref{sec:imcontsub} has been applied to the data. Note that for the L2 data the procedure has been applied to the full frequency range prior to generating the sub-cubes, in order to avoid any discontinuities at the sub-cube boundaries.}

\end{itemize}

The per-channel circular Gaussian beam sizes for the homogenised or cleaned cubes (including the continuum-subtracted set) are provided in a table stored as an additional header data unit (index 1) in the FITS file, as per the {\sc casa} multiple beams table scheme.

\begin{table*}
\begin{minipage}{176mm}
\centering
\caption{Properties of the image products that form the MIGHTEE-\HI~Data Release 1.}
\begin{tabular}{lllllll} \hline
Filename   & $\nu_{\mathrm{min}}$ [MHz] & $\nu_{\mathrm{max}}$ [MHz] & $\Delta_{\nu}$ [kHz] & N$_{\mathrm{chans}}$ & $z_{\mathrm{HI}}^{\mathrm{max}}$ & $z_{\mathrm{OH}}^{\mathrm{max}}$ \\ \hline
MIGHTEE-HI\_DR1\_COSMOS\_L1\_r0p0\_conv\_0000-1817.fits & 960.14 & 1150.13 & 104.5 & 1818 & 0.48 & 0.74 \\
MIGHTEE-HI\_DR1\_COSMOS\_L1\_r0p0\_conv\_contsub\_0000-1817.fits & 960.14 & 1150.13 & 104.5 & 1818 & 0.48 & 0.74 \\
MIGHTEE-HI\_DR1\_COSMOS\_L1\_r0p0\_maser\_clean\_conv.fits & 965.46 & 986.25 & 104.5 & 199 & 0.47 & 0.73 \\
MIGHTEE-HI\_DR1\_COSMOS\_L1\_r0p0\_maser\_clean\_conv\_contsub.fits & 965.46 & 986.25 & 104.5 & 199 & 0.47 & 0.73 \\
MIGHTEE-HI\_DR1\_COSMOS\_L1\_r0p5\_conv\_0000-1817.fits & 960.14 & 1150.13 & 104.5 & 1818 & 0.48 & 0.74 \\
MIGHTEE-HI\_DR1\_COSMOS\_L1\_r0p5\_conv\_contsub\_0000-1817.fits & 960.14 & 1150.13 & 104.5 & 1818 & 0.48 & 0.74 \\
MIGHTEE-HI\_DR1\_COSMOS\_L1\_r0p5\_dirty\_0000-1817.fits & 960.14 & 1150.13 & 104.5 & 1818 & 0.48 & 0.74 \\
MIGHTEE-HI\_DR1\_COSMOS\_L1\_r0p5\_dirty\_contsub\_0000-1817.fits & 960.14 & 1150.13 & 104.5 & 1818 & 0.48 & 0.74 \\
MIGHTEE-HI\_DR1\_COSMOS\_L2\_r0p0\_clean\_conv\_0001-1055.fits & 1290.15 & 1317.71 & 26.13 & 1055 & 0.1 & 0.29 \\
MIGHTEE-HI\_DR1\_COSMOS\_L2\_r0p0\_clean\_conv\_1001-2055.fits & 1316.27 & 1343.83 & 26.13 & 1055 & 0.08 & 0.27 \\
MIGHTEE-HI\_DR1\_COSMOS\_L2\_r0p0\_clean\_conv\_2001-3055.fits & 1342.4 & 1369.96 & 26.13 & 1055 & 0.06 & 0.24 \\
MIGHTEE-HI\_DR1\_COSMOS\_L2\_r0p0\_clean\_conv\_3001-4055.fits & 1368.52 & 1396.08 & 26.13 & 1055 & 0.04 & 0.22 \\
MIGHTEE-HI\_DR1\_COSMOS\_L2\_r0p0\_clean\_conv\_4001-5056.fits & 1394.65 & 1422.21 & 26.13 & 1055 & 0.02 & 0.2 \\
MIGHTEE-HI\_DR1\_COSMOS\_L2\_r0p0\_clean\_conv\_5002-6056.fits & 1420.77 & 1448.34 & 26.13 & 1055 & N/A & 0.17 \\
MIGHTEE-HI\_DR1\_COSMOS\_L2\_r0p0\_clean\_conv\_6002-7056.fits & 1446.9 & 1474.46 & 26.13 & 1055 & N/A & 0.15 \\
MIGHTEE-HI\_DR1\_COSMOS\_L2\_r0p0\_clean\_conv\_7002-8056.fits & 1473.02 & 1500.59 & 26.13 & 1055 & N/A & 0.13 \\
MIGHTEE-HI\_DR1\_COSMOS\_L2\_r0p0\_clean\_conv\_8002-8802.fits & 1499.15 & 1520.08 & 26.13 & 801 & N/A & 0.11 \\
MIGHTEE-HI\_DR1\_COSMOS\_L2\_r0p0\_clean\_conv\_contsub\_0001-1055.fits & 1290.15 & 1317.71 & 26.13 & 1055 & 0.1 & 0.29 \\
MIGHTEE-HI\_DR1\_COSMOS\_L2\_r0p0\_clean\_conv\_contsub\_1001-2055.fits & 1316.27 & 1343.83 & 26.13 & 1055 & 0.08 & 0.27 \\
MIGHTEE-HI\_DR1\_COSMOS\_L2\_r0p0\_clean\_conv\_contsub\_2001-3055.fits & 1342.4 & 1369.96 & 26.13 & 1055 & 0.06 & 0.24 \\
MIGHTEE-HI\_DR1\_COSMOS\_L2\_r0p0\_clean\_conv\_contsub\_3001-4055.fits & 1368.52 & 1396.08 & 26.13 & 1055 & 0.04 & 0.22 \\
MIGHTEE-HI\_DR1\_COSMOS\_L2\_r0p0\_clean\_conv\_contsub\_4001-5056.fits & 1394.65 & 1422.21 & 26.13 & 1055 & 0.02 & 0.2 \\
MIGHTEE-HI\_DR1\_COSMOS\_L2\_r0p0\_clean\_conv\_contsub\_5002-6056.fits & 1420.77 & 1448.34 & 26.13 & 1055 & N/A & 0.17 \\
MIGHTEE-HI\_DR1\_COSMOS\_L2\_r0p0\_clean\_conv\_contsub\_6002-7057.fits & 1446.9 & 1474.46 & 26.13 & 1055 & N/A & 0.15 \\
MIGHTEE-HI\_DR1\_COSMOS\_L2\_r0p0\_clean\_conv\_contsub\_7002-8056.fits & 1473.02 & 1500.59 & 26.13 & 1055 & N/A & 0.13 \\
MIGHTEE-HI\_DR1\_COSMOS\_L2\_r0p0\_clean\_conv\_contsub\_8002-8802.fits & 1499.15 & 1520.08 & 26.13 & 801 & N/A & 0.11 \\
MIGHTEE-HI\_DR1\_COSMOS\_L2\_r0p0\_dirty\_0001-1055.fits & 1290.15 & 1317.71 & 26.13 & 1055 & 0.1 & 0.29 \\
MIGHTEE-HI\_DR1\_COSMOS\_L2\_r0p0\_dirty\_1001-2055.fits & 1316.27 & 1343.83 & 26.13 & 1055 & 0.08 & 0.27 \\
MIGHTEE-HI\_DR1\_COSMOS\_L2\_r0p0\_dirty\_2001-3055.fits & 1342.4 & 1369.96 & 26.13 & 1055 & 0.06 & 0.24 \\
MIGHTEE-HI\_DR1\_COSMOS\_L2\_r0p0\_dirty\_3001-4055.fits & 1368.52 & 1396.08 & 26.13 & 1055 & 0.04 & 0.22 \\
MIGHTEE-HI\_DR1\_COSMOS\_L2\_r0p0\_dirty\_4001-5055.fits & 1394.65 & 1422.21 & 26.13 & 1055 & 0.02 & 0.2 \\
MIGHTEE-HI\_DR1\_COSMOS\_L2\_r0p0\_dirty\_5001-6055.fits & 1420.77 & 1448.34 & 26.13 & 1055 & N/A & 0.17 \\
MIGHTEE-HI\_DR1\_COSMOS\_L2\_r0p0\_dirty\_6001-7055.fits & 1446.9 & 1474.46 & 26.13 & 1055 & N/A & 0.15 \\
MIGHTEE-HI\_DR1\_COSMOS\_L2\_r0p0\_dirty\_7001-8055.fits & 1473.02 & 1500.59 & 26.13 & 1055 & N/A & 0.13 \\
MIGHTEE-HI\_DR1\_COSMOS\_L2\_r0p0\_dirty\_8001-8802.fits & 1499.15 & 1520.1 & 26.13 & 802 & N/A & 0.11 \\
MIGHTEE-HI\_DR1\_COSMOS\_L2\_r0p5\_clean\_conv\_0001-1055.fits & 1290.15 & 1317.71 & 26.13 & 1055 & 0.1 & 0.29 \\
MIGHTEE-HI\_DR1\_COSMOS\_L2\_r0p5\_clean\_conv\_1001-2055.fits & 1316.27 & 1343.83 & 26.13 & 1055 & 0.08 & 0.27 \\
MIGHTEE-HI\_DR1\_COSMOS\_L2\_r0p5\_clean\_conv\_2001-3055.fits & 1342.4 & 1369.96 & 26.13 & 1055 & 0.06 & 0.24 \\
MIGHTEE-HI\_DR1\_COSMOS\_L2\_r0p5\_clean\_conv\_3001-4055.fits & 1368.52 & 1396.08 & 26.13 & 1055 & 0.04 & 0.22 \\
MIGHTEE-HI\_DR1\_COSMOS\_L2\_r0p5\_clean\_conv\_4001-5055.fits & 1394.65 & 1422.21 & 26.13 & 1055 & 0.02 & 0.2 \\
MIGHTEE-HI\_DR1\_COSMOS\_L2\_r0p5\_clean\_conv\_5001-6055.fits & 1420.77 & 1448.34 & 26.13 & 1055 & N/A & 0.17 \\
MIGHTEE-HI\_DR1\_COSMOS\_L2\_r0p5\_clean\_conv\_6001-7055.fits & 1446.9 & 1474.46 & 26.13 & 1055 & N/A & 0.15 \\
MIGHTEE-HI\_DR1\_COSMOS\_L2\_r0p5\_clean\_conv\_7001-8055.fits & 1473.02 & 1500.59 & 26.13 & 1055 & N/A & 0.13 \\
MIGHTEE-HI\_DR1\_COSMOS\_L2\_r0p5\_clean\_conv\_8001-8802.fits & 1499.15 & 1520.1 & 26.13 & 802 & N/A & 0.11 \\
MIGHTEE-HI\_DR1\_COSMOS\_L2\_r0p5\_clean\_conv\_contsub\_0001-1055.fits & 1290.15 & 1317.71 & 26.13 & 1055 & 0.1 & 0.29 \\
MIGHTEE-HI\_DR1\_COSMOS\_L2\_r0p5\_clean\_conv\_contsub\_1001-2055.fits & 1316.27 & 1343.83 & 26.13 & 1055 & 0.08 & 0.27 \\
MIGHTEE-HI\_DR1\_COSMOS\_L2\_r0p5\_clean\_conv\_contsub\_2001-3055.fits & 1342.4 & 1369.96 & 26.13 & 1055 & 0.06 & 0.24 \\
MIGHTEE-HI\_DR1\_COSMOS\_L2\_r0p5\_clean\_conv\_contsub\_3001-4055.fits & 1368.52 & 1396.08 & 26.13 & 1055 & 0.04 & 0.22 \\
MIGHTEE-HI\_DR1\_COSMOS\_L2\_r0p5\_clean\_conv\_contsub\_4001-5055.fits & 1394.65 & 1422.21 & 26.13 & 1055 & 0.02 & 0.2 \\
MIGHTEE-HI\_DR1\_COSMOS\_L2\_r0p5\_clean\_conv\_contsub\_5001-6055.fits & 1420.77 & 1448.34 & 26.13 & 1055 & N/A & 0.17 \\
MIGHTEE-HI\_DR1\_COSMOS\_L2\_r0p5\_clean\_conv\_contsub\_6001-7055.fits & 1446.9 & 1474.46 & 26.13 & 1055 & N/A & 0.15 \\
MIGHTEE-HI\_DR1\_COSMOS\_L2\_r0p5\_clean\_conv\_contsub\_7001-8055.fits & 1473.02 & 1500.59 & 26.13 & 1055 & N/A & 0.13 \\
MIGHTEE-HI\_DR1\_COSMOS\_L2\_r0p5\_clean\_conv\_contsub\_8001-8802.fits & 1499.15 & 1520.1 & 26.13 & 802 & N/A & 0.11 \\
MIGHTEE-HI\_DR1\_COSMOS\_L2\_r0p5\_dirty\_0000-1054.fits & 1290.12 & 1317.68 & 26.13 & 1055 & 0.1 & 0.29 \\
MIGHTEE-HI\_DR1\_COSMOS\_L2\_r0p5\_dirty\_1000-2054.fits & 1316.25 & 1343.81 & 26.13 & 1055 & 0.08 & 0.27 \\
MIGHTEE-HI\_DR1\_COSMOS\_L2\_r0p5\_dirty\_2000-3054.fits & 1342.37 & 1369.93 & 26.13 & 1055 & 0.06 & 0.24 \\
MIGHTEE-HI\_DR1\_COSMOS\_L2\_r0p5\_dirty\_3000-4054.fits & 1368.5 & 1396.06 & 26.13 & 1055 & 0.04 & 0.22 \\
MIGHTEE-HI\_DR1\_COSMOS\_L2\_r0p5\_dirty\_4000-5054.fits & 1394.62 & 1422.18 & 26.13 & 1055 & 0.02 & 0.2 \\
MIGHTEE-HI\_DR1\_COSMOS\_L2\_r0p5\_dirty\_5000-6054.fits & 1420.75 & 1448.31 & 26.13 & 1055 & N/A & 0.17 \\
MIGHTEE-HI\_DR1\_COSMOS\_L2\_r0p5\_dirty\_6000-7054.fits & 1446.87 & 1474.43 & 26.13 & 1055 & N/A & 0.15 \\
MIGHTEE-HI\_DR1\_COSMOS\_L2\_r0p5\_dirty\_7000-8054.fits & 1473.0 & 1500.56 & 26.13 & 1055 & N/A & 0.13 \\
MIGHTEE-HI\_DR1\_COSMOS\_L2\_r0p5\_dirty\_8000-8804.fits & 1499.12 & 1520.15 & 26.13 & 805 & N/A & 0.11 \\
\hline
\end{tabular}
\label{tab:data}
\end{minipage}
\end{table*}

\addtocounter{table}{-1}

\begin{table*}
\begin{minipage}{176mm}
\centering
\caption{\emph{(continued)} Properties of the image products that form the MIGHTEE-\HI~Data Release 1.}
\begin{tabular}{lllllll} \hline
Filename   & $\nu_{\mathrm{min}}$ [MHz] & $\nu_{\mathrm{max}}$ [MHz] & $\Delta_{\nu}$ [kHz] & N$_{\mathrm{chans}}$ & $z_{\mathrm{HI}}^{\mathrm{max}}$ & $z_{\mathrm{OH}}^{\mathrm{max}}$ \\ \hline
MIGHTEE-HI\_DR1\_COSMOS\_L2\_r1p0\_clean\_conv\_0001-1055.fits & 1290.12 & 1317.68 & 26.13 & 1055 & 0.1 & 0.29 \\
MIGHTEE-HI\_DR1\_COSMOS\_L2\_r1p0\_clean\_conv\_1001-2055.fits & 1316.25 & 1343.81 & 26.13 & 1055 & 0.08 & 0.27 \\
MIGHTEE-HI\_DR1\_COSMOS\_L2\_r1p0\_clean\_conv\_2001-3055.fits & 1342.37 & 1369.93 & 26.13 & 1055 & 0.06 & 0.24 \\
MIGHTEE-HI\_DR1\_COSMOS\_L2\_r1p0\_clean\_conv\_3001-4055.fits & 1368.5 & 1396.06 & 26.13 & 1055 & 0.04 & 0.22 \\
MIGHTEE-HI\_DR1\_COSMOS\_L2\_r1p0\_clean\_conv\_4001-5055.fits & 1394.62 & 1422.18 & 26.13 & 1055 & 0.02 & 0.2 \\
MIGHTEE-HI\_DR1\_COSMOS\_L2\_r1p0\_clean\_conv\_5001-6055.fits & 1420.75 & 1448.31 & 26.13 & 1055 & N/A & 0.17 \\
MIGHTEE-HI\_DR1\_COSMOS\_L2\_r1p0\_clean\_conv\_6001-7055.fits & 1446.87 & 1474.43 & 26.13 & 1055 & N/A & 0.15 \\
MIGHTEE-HI\_DR1\_COSMOS\_L2\_r1p0\_clean\_conv\_7001-8055.fits & 1473.0 & 1500.56 & 26.13 & 1055 & N/A & 0.13 \\
MIGHTEE-HI\_DR1\_COSMOS\_L2\_r1p0\_clean\_conv\_8001-8803.fits & 1499.12 & 1520.1 & 26.13 & 803 & N/A & 0.11 \\
MIGHTEE-HI\_DR1\_COSMOS\_L2\_r1p0\_clean\_conv\_contsub\_0001-1055.fits & 1290.12 & 1317.68 & 26.13 & 1055 & 0.1 & 0.29 \\
MIGHTEE-HI\_DR1\_COSMOS\_L2\_r1p0\_clean\_conv\_contsub\_1001-2055.fits & 1316.25 & 1343.81 & 26.13 & 1055 & 0.08 & 0.27 \\
MIGHTEE-HI\_DR1\_COSMOS\_L2\_r1p0\_clean\_conv\_contsub\_2001-3055.fits & 1342.37 & 1369.93 & 26.13 & 1055 & 0.06 & 0.24 \\
MIGHTEE-HI\_DR1\_COSMOS\_L2\_r1p0\_clean\_conv\_contsub\_3001-4055.fits & 1368.5 & 1396.06 & 26.13 & 1055 & 0.04 & 0.22 \\
MIGHTEE-HI\_DR1\_COSMOS\_L2\_r1p0\_clean\_conv\_contsub\_4001-5055.fits & 1394.62 & 1422.18 & 26.13 & 1055 & 0.02 & 0.2 \\
MIGHTEE-HI\_DR1\_COSMOS\_L2\_r1p0\_clean\_conv\_contsub\_5001-6055.fits & 1420.75 & 1448.31 & 26.13 & 1055 & N/A & 0.17 \\
MIGHTEE-HI\_DR1\_COSMOS\_L2\_r1p0\_clean\_conv\_contsub\_6001-7055.fits & 1446.87 & 1474.43 & 26.13 & 1055 & N/A & 0.15 \\
MIGHTEE-HI\_DR1\_COSMOS\_L2\_r1p0\_clean\_conv\_contsub\_7001-8055.fits & 1473.0 & 1500.56 & 26.13 & 1055 & N/A & 0.13 \\
MIGHTEE-HI\_DR1\_COSMOS\_L2\_r1p0\_clean\_conv\_contsub\_8001-8803.fits & 1499.12 & 1520.1 & 26.13 & 803 & N/A & 0.11 \\
MIGHTEE-HI\_DR1\_COSMOS\_L2\_r1p0\_dirty\_0001-1055.fits & 1290.12 & 1317.68 & 26.13 & 1055 & 0.1 & 0.29 \\
MIGHTEE-HI\_DR1\_COSMOS\_L2\_r1p0\_dirty\_1001-2055.fits & 1316.25 & 1343.81 & 26.13 & 1055 & 0.08 & 0.27 \\
MIGHTEE-HI\_DR1\_COSMOS\_L2\_r1p0\_dirty\_2001-3055.fits & 1342.37 & 1369.93 & 26.13 & 1055 & 0.06 & 0.24 \\
MIGHTEE-HI\_DR1\_COSMOS\_L2\_r1p0\_dirty\_3001-4055.fits & 1368.5 & 1396.06 & 26.13 & 1055 & 0.04 & 0.22 \\
MIGHTEE-HI\_DR1\_COSMOS\_L2\_r1p0\_dirty\_4001-5055.fits & 1394.62 & 1422.18 & 26.13 & 1055 & 0.02 & 0.2 \\
MIGHTEE-HI\_DR1\_COSMOS\_L2\_r1p0\_dirty\_5001-6055.fits & 1420.75 & 1448.31 & 26.13 & 1055 & N/A & 0.17 \\
MIGHTEE-HI\_DR1\_COSMOS\_L2\_r1p0\_dirty\_6001-7055.fits & 1446.87 & 1474.43 & 26.13 & 1055 & N/A & 0.15 \\
MIGHTEE-HI\_DR1\_COSMOS\_L2\_r1p0\_dirty\_7001-8055.fits & 1473.0 & 1500.56 & 26.13 & 1055 & N/A & 0.13 \\
MIGHTEE-HI\_DR1\_COSMOS\_L2\_r1p0\_dirty\_8001-8803.fits & 1499.12 & 1520.1 & 26.13 & 803 & N/A & 0.11 \\
\hline
\end{tabular}
\label{tab:data}
\end{minipage}
\end{table*}


\bsp    
\label{lastpage}
\end{document}